%% file: main.tex
\documentclass[twoside,11pt]{article}

\usepackage[abbrvbib, preprint]{jmlr2e}

\input{mathcommands.tex}

\usepackage{enumerate,bbm}
\usepackage{booktabs}
\usepackage[caption=false]{subfig}
\usepackage[toc,page,title,titletoc,header]{appendix}
\usepackage{algorithm,algpseudocode}
\usepackage{multirow}
\usepackage{color}

\ShortHeadings{Distributed Learning of Finite Gaussian Mixtures}{Zhang and Chen}
\firstpageno{1}

\begin{document}

\title{Distributed Learning of Finite Gaussian Mixtures}

\author{\name Qiong Zhang \email qiong.zhang@stat.ubc.ca \\
       \addr Department of Statistics\\
       University of British Columbia\\
       Vancouver, BC V6T 1Z4, Canada
       \AND
       \name Jiahua Chen \email jhchen@stat.ubc.ca \\
       \addr Department of Statistics\\
       University of British Columbia\\
       Vancouver, BC V6T 1Z4, Canada}

\editor{Kevin Murphy and Bernhard Sch{\"o}lkopf}

\maketitle

\begin{abstract}
Advances in information technology have led to extremely large datasets that are often kept in different storage centers.
Existing statistical methods must be adapted to overcome the resulting computational obstacles while retaining statistical validity and efficiency. 
Split-and-conquer approaches have been applied in many areas, including quantile processes, regression analysis, principal eigenspaces, and exponential families.
We study split-and-conquer approaches for the distributed learning of finite Gaussian mixtures. 
We recommend a reduction strategy and develop an effective MM algorithm. 
The new estimator is shown to be consistent and retains root-n consistency under some general conditions. 
Experiments based on simulated and real-world data show that the proposed split-and-conquer approach has comparable statistical performance with the global estimator based on the full dataset, if the latter is feasible. 
It can even outperform the global estimator if the model assumption does not match the real-world data.
It also has better statistical and computational performance than some existing methods.
\end{abstract}


\section{Introduction}
In the era of big data, the sizes of the datasets for various applications may be so large that they cannot be stored on a single machine. 
For instance, according to~\citet{corbett2013spanner}, Google distributes its huge database around the world.
Distributed data storage is also natural when the datasets are collected and managed by independent agencies.
Examples include patient information collected from different hospitals and data collected by different government agencies~\citep{agrawal2003information}.
Privacy considerations may make it difficult or even impossible to pool the separate collections of data into a single dataset stored on a single facility. 
If the dataset is stored on a single machine, it may not be possible to load all of it into the computer memory. 
Data analysis methods should therefore be designed so that they can work with subsets of the dataset, in parallel or sequentially. 
The information extracted from the subsets can then be combined to enable conclusions about the whole population. 
In this context, a two-step split-and-conquer procedure is often used: 
\begin{enumerate}[(i)]
\item 
Local inference: standard inference is carried out on local machines;
\item 
Aggregation: the local results are transmitted to a central machine and aggregated. 
\end{enumerate}
The split-and-conquer approaches address privacy concerns by sharing only summary statistics across machines.
This also avoids a potentially high transmission cost that may exceed the computational cost~\citep{jaggi2014communication}.

There are many split-and-conquer approaches. 
Examples include~\citet{chen2014split} on generalized linear models,~\citet{zhang2015divide} on the kernel ridge regression model,~\citet{battey2015distributed} on high-dimensional generalizations of the Wald and Score tests under generalized linear models,~\citet{chang2017divide} on local average regression models, and~\citet{fan2019distributed} on principal eigenspaces.
These methods first obtain multiple local estimates of the model parameters based on the local datasets.
They may then perform a linear averaging to aggregate these estimates.

We are interested in the finite mixture model, and its parameter space has discrete distributions with a finite number of support points.
A simple average of the local estimates leads to a mixture with redundant and spurious subpopulations.
An interesting research problem therefore arises for such models: \emph{how to develop an aggregation approach that is sensitive and computationally efficient with the least distortion?}

Finite mixture models provide a natural representation of heterogeneity in various populations~\citep{pearson1894contributions}. 
They have been widely used, for example in image generation~\citep{kolouri2017sliced}, image segmentation~\citep{farnoosh2008image}, object tracking~\citep{santosh2013tracking}, and signal processing~\citep{kostantinos2000gaussian}.
The flexibility of finite mixture models makes them ideal for distributions of unknown shape. 
These models are usually learned via the maximum likelihood estimate (MLE) through a well-established EM-algorithm~\citep{wu1983convergence}. 
The EM-algorithm, although easy to implement, is notoriously slow and may fail to locate the global maximum of the likelihood.
Researchers have devoted much energy to improving the implementation of the EM 
algorithm~\citep{meilijson1989fast,meng1993maximum,liu1994ecme}. 
Distributed versions of this algorithm have been proposed in~\citet{nowak2003distributed},~\citet{safarinejadian2010distributed}, and~\citet{chen2013parallel}. 
These approaches are expensive because at each iteration they communicate
summary statistics across the local machines.

In this paper, we develop a novel split-and-conquer procedure for a finite Gaussian mixture.
We employ the penalized MLE of the finite mixture model to obtain local estimates of the mixing distribution, and we investigate a number of aggregation approaches.
We find that a specific reduction approach has satisfactory statistical efficiency at a reasonable computational cost.
This method first pools the local estimates to form a mixture model with an excessive number of subpopulations. It then searches for a mixture with a designated order that is optimal according to some divergence criteria. 
An MM algorithm is developed for the numerical computation.
We choose a divergence based on the transportation distance so that the corresponding estimator retains statistical efficiency and computational simplicity.

The remainder of the paper is structured as follows.
In Section~\ref{sec:background}, we briefly discuss finite Gaussian mixture models (GMMs), the composite transportation distance, and some related literature. 
In Section~\ref{sec:method}, we study split-and-conquer methods and recommend a specific one.
In Section~\ref{sec:algorithm}, we describe the algorithm for computing the recommended estimator.
In Section~\ref{sec:theory}, we show that this estimator has the best possible statistical convergence rate under certain conditions. 
Numerical experiments on simulated and real data are presented in Section~\ref{sec:experiment}, and our approach performs well.
In Section~\ref{sec:conclusion}, we provide a discussion and some concluding remarks. 
The technical details are included in the appendix.

\section{Related Work}
Learning of the mixture is the most fundamental task in modeling the data with mixture models. 
In this section, we review the related methods for learning finite mixture models under distributed datasets.

\subsection{Distributed EM Algorithm}
The EM algorithm is the most widely used numerical algorithm for the numerical computation of MLE. 
When the datasets are stored in a distributed fashion, the corresponding distributed version of the EM algorithm is developed in~\citet{nowak2003distributed},~\citet{safarinejadian2010distributed}, and~\citet{chen2013parallel}.
In the E-step of the EM algorithm, the conditional expectations
can be computed locally on each local machine. 
To carry out the M-step based on the full dataset, the local summary statistics need to be sent to the central machine to be aggregated. 
A na\"ive implementation of the EM algorithm under distributed learning thus requires $2M$ transmission of summary statistics each of dimension $3d$ at each iteration, which is expensive.
To reduce the transmission cost,~\citet{nowak2003distributed} proposes to perform incremental E and M steps~\citep{neal1998view} at each local machine. 
At each iteration, each local machine receives the summary statistics of dimension $3d$ from previous local machine, compute the conditional expectation and perform incremental M-step locally, then send the updated summary statistics to the next local machine to run the next iteration. 
Although the rounds of transmission is reduced in this approach, statistics of dimension $3d$ still needs to be transmitted at each iteration. 
Since the transmission cost may be more expensive than the computational cost under distributed learning~\citep{jaggi2014communication}, this approach is therefore more expensive than our proposed approach since we only require one round of communication during the whole process.
Similar issue exists for~\citet{safarinejadian2010distributed} and~\citet{chen2013parallel}.

\subsection{Learning at Scale via Coresets}
A coreset is defined as a weighted sample of the full dataset so that the coreset is representative of the full dataset in certain sense. 
As the size of the coreset is much smaller compared to the full dataset, fitting a model on the coreset $\mathcal{C}$ is computationally much faster than fitting the model on the full dataset $\mathcal{X}$.
To learn finite mixture models on large scale datasets,~\citet{feldman2011scalable} and~\citet{lucic2017training} considers how to construct the coreset when data are IID samples from finite Gaussian mixtures.
To construct the coreset, they subsample observations from the dataset.
When the observations are random samples from a mixture, due to the value of the mixing weights, the number of the observations from each subpopulation can be quite imbalanced. 
In such cases, if observations are sampled uniformly at random, then it is very likely that only observations in the subpopulation with high mixing weight are included in the subsample.
To avoid this issue,~\citet{lucic2017training} proposes to build a coreset via importance sampling to approximate a normalized log-likelihood. 
Their goal is that the relative difference in the normalized log-likelihood based on the full dataset and the normalized log-likelihood based on the coreset is smaller than some small value $\epsilon$.
The normalized log-likelihood based on the coreset can be viewed as an estimator for the normalized log-likelihood, whose variance depends on the sampling distribution in the importance sampling. 
They find the sampling distribution under the mixture model so that the variance of the estimated normalized log-likelihood is minimized. 
Surprisingly,~\citet{lucic2017training} find out that when the parameter space is compact, they are able to construct a coreset with size $|\mathcal{C}| = O(d^4K^6\epsilon^{-2})$ where $\epsilon$ is the precision of the approximation of the normalized log-likelihood.
It can be seen that the coreset size does not depend on the size of the full dataset. 

Under the distributed setting, they propose to build a coreset on each local machine. These local coresets are then merged as follows: a coreset is constructed for two pairs of coresets and continue until there is a single coreset. This process takes $O(\log M)$ merges.
Then the model is learned based on the final merged coreset via the MLE. 
We compare our method to this approach both in the simulation and the real dataset. 
We found in our experiment that a large value of regularization $\lambda$ is required to make the covariance positive definite. 
In our experiment, we learn the mixture on the merged coreset via pMLE.

\subsection{Other Approaches}

\noindent
\textbf{Bayesian Moment Matching} \cite{jaini2016online} proposes a Bayesian approach for learning Gaussian mixture for streaming data. 
They assume the order of the mixture is known and the posterior is updated recursively by Bayes' theorem as new data comes in.
The model is learned by the posterior mean after observing all data.
Despite their efforts to approximate the posterior mixture under the online learning setting, the authors fail to notice the problem of multidomality and non-identifiablility with Bayesian inference in finite mixture models.
Since the posterior mean under finite mixture does not have a closed form, random samples are usually drawn from the posterior and the sample mean is used to approximate the posterior mean. 
Due to the label switching problem, we can arbitrarily permute the parameter without affecting the likelihood.
As a result, we are not able to identify from which posterior distribution these random samples are generated from, but to get a sensible estimator for the posterior mean of $\mu_k$ for example, the samples need to be generated from the $\mu_k|\gX$.
See~\citet{jasra2005markov} and~\citet[Section 23.2.3.1]{murphy2012machine}.
Moreover, the posterior distribution given the full dataset by combining the local posterior in (10) of~\citet{jaini2016online} is not a well defined probability distribution.

\noindent
\textbf{ADMM for Distributed Optimization} The Alternating Direction Method of Multipliers (ADMM) is widely used approach for distributed optimization~\citep{boyd2011distributed}.
The distributed learning is formulated as an optimization problem
\[
\min \sum_{m=1}^{M} f(\theta_m|\gX_{m})\quad\text{subject to } \theta_m-\theta=0,~\forall m=1,2,\ldots, M.
\]
In our case, $f(\theta_m|\gX_m)$ is the penalized log-likelihood function based on the observation on the $m$th local machine.
At each iteration of the ADMM algorithm, we have
\begin{equation}
\label{eq:admm_update}
\begin{split}
&\theta_{m}^{t+1} = \argmin_{\theta_m} \{f(\theta_m|\gX_m) + (\eta_m^{t})^{\tau}(\theta_m-\bar{\theta}^{t}) + \rho/2\|\theta_m-\bar{\theta}^{t}\|_2^2\}\\
&\eta_m^{t+1} = \eta_m^{t} + \rho(\theta_m^{t+1} - \bar{\theta}^{t+1})
\end{split}
\end{equation}
First of all, similar to the distributed EM algorithms, ADMM requires the local machines communicate with the central machine at each iteration.
Under the distributed learning framework, the communication cost could be more expensive than the computational cost~\citep{jaggi2014communication}.
Moreover, as given in~\eqref{eq:admm_update}, at each step of the iteration of the ADMM, the average of local outputs $\bar{\theta}^{t}$ needs to be computed, which is not sensible under finite mixture model due to the label switching problem.
Otherwise, we could perform local inference and directly average the local estimates for aggregation as is the case when the parameter space is Euclidean.
\section{Background}
\label{sec:background}

In this section, we define finite mixture models, present the EM-algorithm, and introduce the concept of distance in the space of (mixing and mixture) distributions.

\subsection{Finite Mixture Models and Model-based Clustering}
\label{sec:GMM}

A statistical model is a family of distributions. 
In some applications, we may regard a dataset as containing observed values of a random variable with a distribution in a parametric family such as the Gaussian distribution.
To explain heterogeneity in the data, we may suggest that the population is made of several subpopulations so that samples from different subpopulations have distributions in the same family but with different parameter values.
If the subpopulation memberships of the individual units are unknown to us, the
data form a sample from a mixture distribution.
Mathematically, a finite mixture model is given as follows.

\begin{definition}[Finite Mixture Model]
Let $\mathcal{F} = \{ f(\cdot |\theta): \theta \in \Theta\}$ be a parametric family of density functions with respect to some $\sigma$-finite measure.
Let $G= \sum_{k=1}^{K} w_k \delta_{\theta_k}$ be a discrete probability measure, assigning probability $w_k$ to parameter value $\theta_k$, for some integer $K>0$ on $\Theta$. 
The finite mixture distribution of $\mathcal{F}$ with mixing distribution $G$ has density function
\begin{equation*}
\label{def:general_mixture_density}
    f(x | G) := \int f(x|\theta)dG(\theta) = \sum_{k=1}^K w_k f(x|\theta_k).
\end{equation*}
\end{definition}

In this definition, $\btheta = (\theta_1, \ldots, \theta_K)^\tau$ is a vector of subpopulation parameters, and the $f(\cdot |\theta_k)$s are called subpopulation or component density functions.
Let the $K$-dimensional simplex be
\[
\Delta_{K}= \{(w_1, \ldots, w_K): w_i \geq 0, \sum_{i=1}^K w_i=1\}.
\]
The components of the vector $\bw = (w_1,w_2,\ldots,w_K)^{\tau} \in \Delta_{K}$ are called the mixing weights.
We use $F(x |\theta)$ and $F(x |G)$ respectively for the cumulative distribution functions (CDFs) of $f(x|\theta)$ and $f(x|G)$. 
The space of mixing distributions of order up to $K$ is denoted
\begin{equation}
\label{eq:G_K}
\mathbb{G}_{K}
=
\big \{
G=\sum_{k=1}^K w_k\delta_{\theta_k},  
  \bw \in \Delta_{K},  \btheta \in \Theta^K
\big \}.
\end{equation}
An order $K$ mixture has its mixing distribution in $\mathbb{G}_{K}-\mathbb{G}_{K-1}$.

One of the earliest two-component GMMs was given by~\citet{pearson1894contributions}:
he suspected that the skewness identified in the distribution of the biometrics of $1,000$ crabs was due to the presence of two non-identified species.
There are also many examples of mixture models in financial contexts.
For example, if the stock prices in the hidden periods in ``normal'' or ``extreme'' states have different Gaussian distributions, then their marginal distributions are two-component GMMs~\citep{liesenfeld2001generalized}.
For a detailed description of the theory and applications of the finite mixture models, see~\citet{mclachlan2004finite} and \citet{fruhwirth2006finite}.
Finite mixtures may be based on many different distribution families, but the finite GMM is by far the most widely used.

\begin{example}[Finite Gaussian Mixtures]
Let the density function of a $d$-dimensional Gaussian random variable with mean vector $\bmu$ and covariance matrix $\bSigma$ be 
\[
\phi(\vx | \bmu, \bSigma) 
= \det\{ 2\pi\bSigma\} ^{-1/2}
\exp \big \{-\frac{1}{2}(\vx-\bmu)^\tau \bSigma^{-1}(\vx-\bmu) \big \}
\]
and $\Phi(\vx| \bmu, \bSigma)$ be its CDF. 
A $K$-component GMM is a distribution with density and CDF given by
\[
     \phi(\vx |G)  = \sum_{k=1}^K w_k \phi(\vx| \bmu_k,\bSigma_k); ~~~
     \Phi(\vx |G)  = \sum_{k=1}^K w_k \Phi (\vx| \bmu_k,\bSigma_k).
\]
In this example, $G$ is a discrete distribution in the space of the $d$-dimensional vector $\bmu$ and $d \times d$ positive definite matrix $\boldsymbol\Sigma$.
\end{example}

The most popular learning approach for finite mixture models relies on the maximum likelihood estimate (MLE). 
Because under the GMM the likelihood is unbounded, the MLE of the mixing distribution $G$ is not well defined. 
However, the log-likelihood function can be regularized by the addition of a penalty function.
Then the $G$ that maximizes the penalized likelihood is well defined and consistent~\citep{chen2009inference}.
Specifically, let $\Xcal = \{\vx_1,\vx_2,\ldots,\vx_n\}$ be observed values on a set of independent and identically distributed (IID) random vectors with finite Gaussian mixture distribution $\phi(\bolds x|G)$ of order $K$.  
The log-likelihood function of $G$ is given by
\begin{equation*}
\label{eq:log_likelihood}
\ell_n(G | \Xcal) 
= 
\sum_{i=1}^{n} \log \phi(\vx_i| G) 
= 
\sum_{i=1}^{n} \log 
\big \{
\sum_{k=1}^K w_k\phi\big (\vx_{i}|\bmu_k,\bSigma_k \big )
\big\}.
\end{equation*}
This likelihood function is unbounded: its value goes to infinity for a specific combination of $\bmu_k$ and some degenerating $\bSigma_k$.
Let $S_{x}$ be the sample covariance matrix.~\citet{chen2009inference} recommend the following penalized log-likelihood function
\begin{equation*}
p\ell_n(G | \Xcal) 
= 
\ell_n(G | \Xcal) - a_n\sum_{k} 
\big \{\text{tr}(S_{x} \bSigma_k^{-1}) + \log \text{det}(\bSigma_k) \big \}
\end{equation*}
for some positive $a_n$ that is allowed to depend on $n$, and they learn the mixing distribution $G$ via
\begin{equation}
\label{eq:pMLE}
    \hat{G}_{n}:=\argsup p\ell_n(G| \Xcal).
\end{equation}
The size of $a_n$ controls the strength of the penalty; a recommended value is $n^{-1/2}$. 
Regularizing the likelihood function via a penalty function fixes the problem caused by degenerating $\bSigma_k$.
The pMLE can be shown to be strongly consistent when the number of components has a known upper bound. 
The EM algorithm can be used~\citep{chen2009inference} for the computation.

\subsubsection{Learning at Local Machines}
\label{sec:unsupervised_learning_local}

Suppose we have a random sample $\Xcal = \{\vx_1, \vx_2, \ldots, \vx_N\}$, and it is either randomly partitioned into $M$ subsets stored on $M$ local machines or can be treated as such. 
Let $\Xcal_m$ of size $N_m$ be the dataset on the $m$th local machine.
The local inference learns the mixture via the pMLE
\[
\hat{G}_m 
= 
\argmax \left \{
\sum_{i\in \Xcal_m} 
\log \phi(\vx_i | G) - 
a_{m}
\sum_{k=1}^K \big \{ \text{tr}(S_{m}\bSigma_k^{-1}) + \log\det(\bSigma_k)\big \}
\right\}
\]
under the GMM. We call $\hat G_m$ a \emph{local} estimate.
Here $S_{m}$ is the sample covariance matrix based on the observations in $ \Xcal_m$, and $a_m= N_m^{-1/2}$ as recommended.
The EM-algorithm can readily be adapted to compute the pMLE. 
We present the adapted EM-algorithm for a single dataset $\Xcal=\{\vx_1,\vx_2,\ldots,\vx_n\}$ for which the sample covariance matrix is $S_{x}$.

We first introduce the membership vector $\mathbf{z}_i = (z_{i1}, \ldots, z_{iK})$ for the $i$th unit. 
The $k$th entry of $\mathbf{z}_i$ is $1$ when the response value $\vx_i$
is an observation from the $k$th subpopulation, and $0$ otherwise. 
When the complete data $\{(\mathbf{z}_i, \vx_i), i=1,2,\ldots,n\}$ are available, the penalized likelihood is given by
\[
    p\ell_{n}^c(G) 
    = \sum_{i=1}^{n} \sum_{k=1}^K z_{ik}\log \{ w_k \phi(\vx_i |\bmu_k, \bSigma_k) \}
    -a_n \sum_{k} 
    \big \{ \text{tr}(S_{x}\bSigma_k^{-1} ) + \log \text{det}(\bSigma_k) \big \}.
\]
Given the observed data $\Xcal$ and proposed mixing distribution $G^{(t)}$,
in the E-step we compute the conditional expectation
\[
w_{ik}^{(t)} 
= \mathbb{E}(z_{ik}| \Xcal, G^{(t)})
= \frac{w_k^{(t)} \phi(\vx_i|\bmu_k^{(t)}, \bSigma_k^{(t)})}
           {\sum_{j=1}^{K} w_j^{(t)} \phi(\vx_i| \bmu_j^{(t)},\bSigma_j^{(t)})}.
\]
With this, we define
\begin{align*}
    Q(G;G^{(t)}) 
     =& 
     \sum_{k=1}^K \big  \{  \sum_{i=1}^{n} w_{ik}^{(t)}\big  \} \log w_k 
    - \frac{1}{2} \sum_{k=1}^K  \big  ( 2a_{n} + \sum_{i=1}^{n}  w_{ik}^{(t)}\big  ) \log \text{det} (\bSigma_k) \\
    & 
    - \frac{1}{2} \sum_{k=1}^K \big \{
      2a_{n} \text{tr} \big \{\bSigma_k^{-1}S\big\}+ \sum_{i=1}^{N}w_{ik}^{(t)}(\vx_i - \bmu_k)^\tau
    \bSigma_k^{-1}(\vx_i - \bmu_k) \big \}.
\end{align*}
Note that the subpopulation parameters are well separated in $Q(\cdot; \cdot)$.
The M-step maximizes $Q(G;G^{(t)}) $ with respect to $G$. 
The solution is given by the mixing distribution $G^{(t+1)}$ with
\[
    \begin{cases}
        w_{k}^{(t+1)}   
            &=   n^{-1} \sum_{i=1}^{n} w_{ik}^{(t)}, \\
        \bmu_{k}^{(t+1)}  
            &= \big \{ n w_k^{(t+1)} \big \}^{-1}  \sum_{i=1}^{n} w_{ik}^{(t)} \vx_i, \\
        \bSigma_{k}^{(t+1)}  
            &= \big \{ 2a_{n} + n w_{k}^{(t+1)} \big \}^{-1} \big \{ 2a_{n} S_{x} + S_k^{(t+1)} \big \},
    \end{cases}
\]
where $S_k^{(t+1)} = \sum_{i=1}^{N} w_{ik}^{(t)} (\vx_i - \bmu_k^{(t+1)}) (\vx_i-\bmu_k^{(t+1)} )^\tau$.

The EM-algorithm for the pMLE, like its MLE counterpart, increases the value of the penalized likelihood after each iteration. 
Let $A \geq B$ denote $A-B$ is semi-positive definite.
Because for all $t$, $  \bSigma_{k}^{(t+1)}  \geq \{ 2a_{n}/ (n+2a_{n})\} S_{x} > 0$ with a lower bound not dependent on the parameter values, the penalized
likelihood has a finite upper bound. 
Hence, the above iterative procedure is guaranteed to converge to at least a non-degenerate local maximum.

\subsection{Divergence and Distance on Space of Probability Measures}

To develop split-and-conquer methods for finite mixture models, we need a measure of the closeness between mixing distributions and between mixtures. 
We first introduce the general notions of divergence and distance.

\begin{definition}[Divergence and Distance]
Let $\Theta$ be a space.  A bivariate function $\rho(\cdot, \cdot)$ 
defined on $\Theta$ is a divergence if $\rho(\theta_1, \theta_2) \geq 0$, 
with equality holding if and only if $\theta_1 = \theta_2$.

Suppose that in addition $\rho(\cdot, \cdot)$ satisfies
(i) symmetry, i.e., $\rho(\theta_1,\theta_2) = \rho(\theta_2, \theta_1)$,
and (ii) the triangle inequality, i.e.,
$\rho(\theta_1, \theta_2) \leq \rho(\theta_1, \theta_3) + \rho(\theta_3, \theta_2)$,
for all $\theta_1, \theta_2, \theta_3 \in \Theta$.
Then $\rho(\cdot,\cdot)$ is a distance on $\Theta$.
When $\rho(\cdot, \cdot)$ is a distance, we call $(\Theta, \rho)$ a metric space.
\end{definition}

We need divergences or distances specifically defined on the space of probability measures.
The transportation divergence~\citep{villani2003topics} is particularly useful.
Let $\mathcal{P}(\Theta)$ and $\mathcal{P}(\Theta^2)$ be the spaces of probability measures on $\Theta$ and $\Theta \times \Theta$, respectively, equipped with some compatible $\sigma$-algebras.
For any ${\bpi} \in \mathcal{P}(\Theta^2)$, let its marginal measures
be ${\bpi}_{1,\cdot}$ and ${\bpi}_{\cdot, 2}$.
For any $\eta, \nu \in \mathcal{P}(\Theta)$, we define a space
of distributions on $\Theta \times \Theta$ as follows:
\begin{equation*}
\label{eq:ot_coupling}
\Pi (\eta, \nu) =
\{{\bpi} \in \mathcal{P}(\Theta^2): {\bpi}_{1,\cdot} = \eta,\, {\bpi}_{\cdot,2} = \nu \}.
\end{equation*}
That is, $\Pi (\eta, \nu)$ consists of bivariate distributions for which the marginal
distributions are $\eta$ and $\nu$.
For convenience, we use $\Pi (\eta, \cdot) $
for the space of distributions with first marginal distribution being $\eta$, 
and similarly for $\Pi (\cdot, \nu) $. 
We regard the mixing weights $\bw$ as a distribution on $\Theta$ when appropriate.

\begin{definition}[Transportation Divergence and $r$-Wasserstein distance]
\label{def:Monge-Kantorovich}
Let $c (\cdot, \cdot)$ be a non-negative valued bivariate function on 
$\Theta \times \Theta$ satisfying $c(\theta, \theta) = 0$ for all $\theta \in \Theta$.
The transportation divergence from $\eta$ to $\nu$ is defined to be
\begin{equation*}
\label{eq:ot_divergence}
\mathcal{T}_{c}(\eta,\nu) 
=  \inf \{ \bbE_{\bpi} [ c(X, Y)] : {\bpi} \in \Pi (\eta, \nu) \}.
\end{equation*}
where $\bbE_{\bpi} \{c(X, Y) \}$ is the expectation calculated 
when the joint distribution of $X, Y$ is ${\bpi}$.

If $c(\cdot, \cdot) = D^r(\cdot, \cdot)$ for some $r \geq 1$ and some distance
$D(\cdot, \cdot)$ on $\Theta$, then $\Tcal_c^{1/r}(\cdot, \cdot)$ is called the
$r$-Wasserstein distance with the ground distance $D(\cdot, \cdot)$.
\end{definition}

The joint distribution ${{\bpi}}$ that minimizes $\bbE_{{\bpi}} \{c(X, Y)\}$ is called 
the optimal transportation plan and is usually denoted ${{\bpi}}^*$.
It is the most efficient plan that transports mass distributed 
according to $\eta$ to mass distributed according to $\nu$.

\begin{definition}[Barycenter]
\label{def:barycenter}
Let ($\mathcal{P}(\Theta), \rho$) be a space of probability measures on 
$\Theta$ that is endowed with the divergence $\rho(\cdot, \cdot)$.
Given positive constants $(\lambda_1,\lambda_2,\ldots,\lambda_M) \in \Delta_{M}$,
the (weighted) barycenter of $\nu_1, \ldots, \nu_M \in \mathcal{P}(\Theta)$
is a minimum point of $\sum_{m=1}^M \lambda_m \rho (\nu_m, \eta )$.
\end{definition}

We allow $\rho(\cdot, \cdot)$ to be any divergence function.
When $\rho(\cdot, \cdot)= D^r(\cdot, \cdot)$,
it becomes the usual $r$-Wasserstein barycenter~\citep{cuturi2014fast}.
Conceptually, a barycenter should always be referred to together with information on $\rho$
and the weights $(\lambda_1,\lambda_2,\ldots,\lambda_M)$. However, the formal description
can be tedious and counterproductive. 
We will provide the full information only when necessary. 

\section{Proposed Split-and-Conquer Approach}
\label{sec:method}

Suppose we have an IID random sample $\Xcal = \{\vx_1, \vx_2, \ldots, \vx_N\}$ 
from a parametric distribution $f(\vx|\theta)$.
Suppose that it is partitioned into $M$ subsets $\Xcal_1, \ldots, \Xcal_M$
completely at random and stored on $M$ local machines. 
Let $N_m$ denote the sample size on the $m$th local machine.
Clearly, $\sum_{m=1}^M N_m = N$. 
Let $\hat{\theta}_m$ be the local estimate of $\theta$ based on $\Xcal_m$. 
How should we aggregate the local estimates $\hat{\theta}_m$ in the split-and-conquer framework?

One approach is to combine the local estimates 
by their linear average~\citep{zhang2015divide,chang2017divide}.
The aggregated estimate is then the weighted average
$\bar{\theta} = \sum_{m=1}^M \lambda_m \hat{\theta}_m$
with $\lambda_m$ set to the sample proportion $N_m/N$.
This simple approach is appropriate for parameters in a vector space, but 
it is nonsensical if the average of the parameters is not well defined. 
In the context of finite mixtures, let $\hat{G}_1, \ldots, \hat{G}_m$ be the local estimators.
A natural aggregated estimator of the mixing distribution 
is the weighted average
\begin{equation}
\label{eq:pool_estimator}
\bar G = \sum_{m=1}^M \lambda_m \hat{G}_m.
\end{equation}
Its corresponding mixture has density function
\(
f(x| \bar G) = \sum_{m=1}^M  \lambda_m f(x| \hat G_m)
\).
While $f(x | \bar G)$ is a good estimate of the true mixture
with density function $f(x| G^*)$,  it can be unsatisfactory in terms
of revealing the latent structure of the mixture model.
For instance, this estimator gives a mixture distribution with 
$MK$ subpopulations rather than the assumed $K$, 
which is useless for clustering the dataset into $K$ clusters.

One way to adapt the linear averaging  
is to define a sensible ``average'' in the space of the mixing distributions. 
Recall that the linear average of vectors in the Euclidean space is the centroid (barycenter) 
of these vectors from a geometric point of view. 
Similarly, the local estimators $\hat{G}_1, \ldots, \hat{G}_m$ may be ``averaged'' through their barycenter:
\begin{equation}
\label{estimator:barycenter}
\bar G^C 
= \arginf_{G \in \mathbb{G}_K} \sum_{m=1}^M \lambda_m \rho (\hat G_m , G )
\end{equation}
for some choice of the divergence $\rho(\cdot, \cdot)$,
where $\mathbb{G}_K$ is the space of mixing distributions with $K$ support
points, as defined in~\eqref{eq:G_K}.

The mixing distribution $\bar G$
is likely close to the true mixing distribution $G^*$, 
except for the incorrect number of support points. 
This problem can be solved by approximating $\bar G$ by some $G \in \mathbb{G}_K$.
This suggests another aggregation approach.
Let $\rho(\cdot,\cdot)$ be a divergence in the space of mixing distributions.
We can aggregate the local estimates via the reduction estimator, given by
\begin{equation}
\label{eq:reduction}
\bar{G}^{R} = \arginf_{G\in \mathbb{G}_K} \rho (\bar G,  G).
\end{equation}

These two aggregation approaches, barycenter and reduction, are connected
when specific divergences are used.
They can also be very different.
For example, let the divergence $\rho(\cdot, \cdot)$ in the barycenter definition~\eqref{estimator:barycenter}
or in the reduction estimator~\eqref{eq:reduction} be
the well-known KL-divergence:
\[
\rho(G_1,  G_2) 
=
\dKL( \Phi (\cdot | G_1 ) \| \Phi (\cdot | G_2 ))
=
\int  \phi (x | G_1)  \log\{ \phi (x | G_1) /\phi(x | G_2) \} dx.
\]
In this case, we have
\begin{equation*}
    \begin{split}
        \dKL(\Phi (\cdot| \bar G) \| \Phi (\cdot | G))&=\int  \phi (x| \bar G) \log \big \{\phi (x| \bar G)/ \phi (x | G) \big \} dx =
C_1 - \int  \phi (x| \bar G) \log \phi (x | G) dx \\
&=
C_1 - \sum_{m=1}^M \lambda_m \int  \phi (x| \hat G_m) \log \phi (x | G) dx \\
&=
C_2 + \sum_{m=1}^M \lambda_m  \dKL ( \Phi (\cdot| \hat G_m)\| \Phi (\cdot|G))
    \end{split}
\end{equation*}
where $C_1$ and $C_2$ are constants  not dependent on $G$.
This relationship implies that
\be
\label{eq:equivalentKL}
\bar{G}^{R} 
=
\arginf_{G\in \mathbb{G}_K} \rho (\bar G,  G)
=
\arginf_{G\in \mathbb{G}_K} 
\Big \{ \sum_{m=1}^M \lambda_m   \rho (\hat G_m,  G) \Big \}
=
\bar{G}^{C} .
\ee
Thus, the two aggregation methods give identical aggregated estimators.
\citet{liu2014distributed} consider the use of the KL divergence $\rho(\theta_1,\theta_2)=\dKL(f(\cdot|\theta_1)\|f(\cdot|\theta_2))$ in~\eqref{estimator:barycenter} and hence in~\eqref{eq:reduction} to obtain
fully efficient parameter estimators for exponential family models with
a split-and-conquer approach. 
However, this is not suitable for finite GMMs
for computational reasons. Therefore,~\citet{liu2014distributed} 
find an approximate solution to~\eqref{estimator:barycenter} 
by fitting a $K$-component GMM on
pooled samples of reduced sizes generated from the locally learned GMMs.
Hereafter, we refer to this aggregated estimator as the KL-averaing (KLA) estimator.
It does not retain statistical efficiency under finite mixture, in contrast to
their remarkable achievements for exponential families.

The following example shows that the two methods do not always have the same outcome.

\begin{example}[Barycenter of Two Univariate Two-Component GMMs]
\label{eg:barycenter_of_GMM}
Suppose we wish to aggregate two local estimates given by
\begin{eqnarray*}
\phi(x|G_1) &=&  0.4 \phi(x| -1, 1) + 0.6 \phi(x| 1, 1) :=  0.4 \phi_{-1} + 0.6 \phi_{1},\\
\phi(x|G_2) &=&  0. 6\phi(x| -1, 1) + 0.4 \phi(x| 1, 1) :=  0.6 \phi_{-1} + 0.4 \phi_{1}
\end{eqnarray*}
with $\lambda_1 = \lambda_2 = 0.5$.
We anticipate that whatever distance or divergence we choose, 
the barycenter will be given by
\[
\phi(x|\bar G) =  0.5 \phi(x| -1, 1) + 0.5 \phi(x| 1, 1) = 0.5 \phi_{-1} + 0.5 \phi_1.
\]
Consider the $2$-Wasserstein distance with Euclidean ground distance
between two univariate Gaussian distributions that is given by 
\[
D^2 \big ( \phi(\cdot| \mu_1, \sigma_1^2), \phi(\cdot| \mu_2, \sigma_2^2) \big )
=
(\mu_1 - \mu_2)^2 + (\sigma_1 - \sigma_2)^2.
\]
Surprisingly, we find that the barycenter when $\rho=\bbW_{D, 2}^2(\cdot, \cdot)$ is given by
\[
\phi(x|\bar G^C) =  0.4 \phi(x| -1, 1) + 0.6 \phi(x; 2/3, 1) := 0.4 \phi_{-1} + 0.6 \phi_{2/3} .
\]
\end{example}

We defer the technical details to Appendix~\ref{app:proof_barycenter_of_GMM}.
We create this example extremely sharp to best illustrate this issue. The distortion of barycenter to the level in this example is unlikely in a real world
application. 
The issue here is rooted in the divergence $\rho$ employed in defining the
barycenter; other choices of $\rho$ may lead to a solution consistent with
our intuition. However, for the reduction estimator, whatever divergence $\rho$ is employed, 
we always have $\bar G^R = \bar G$.
This example motivates us to consider the choice of the divergence 
$\rho$ in~\eqref{estimator:barycenter} and~\eqref{eq:reduction}.
We must take into account both statistical efficiency and computational complexity.
For statistical efficiency, the properties of the divergence in~\eqref{estimator:barycenter}
must ensure that the corresponding barycenter matches our intuition. 
Besides, the computation of~\eqref{estimator:barycenter} is more
expensive than that of~\eqref{eq:reduction} given the same $\rho$.
This paper hence starts with investigating the reduction estimator~\eqref{eq:reduction} for split-and-conquer
learning of finite mixtures.


In the machine learning community, approximating a GMM by one with a lower order is called Gaussian 
mixture reduction (GMR)~\citep{williams2006cost,schieferdecker2009gaussian,yu2018density}.
These approaches are usually ad hoc.~\citet{williams2006cost} use an optimization based approach that minimize 
$\rho(G_1,G_2)=L_2(\phi(\cdot|G_1),\phi(\cdot|G_2))$.
Although the $L_2$ distance between two GMMs has an analytical form, 
this optimization is expensive. 
Our key observation is that it is usually difficult to compute the divergence 
between two mixtures, but easy to compute that between two Gaussian distributions. 
This leads to the effective reduction algorithm that we present in the next section. 
For simplicity, we refer to the proposed estimator as the GMR estimator.

\section{Numerical Algorithm for Proposed Approach}
\label{sec:algorithm}

Let $\bar G$ be as defined in~\eqref{eq:pool_estimator}. 
Let the subpopulations in $\bar G$ be $\Phi_i = \Phi(x|\bmu_i, \bSigma_i)$
and the mixing weights be $w_i$ for $i \in [MK]$.
Here $[M]$ is the set $\{1, 2, \ldots, M\}$.
Let $G$ be any mixing distribution of order $K$
with the $K$ subpopulations 
$\Phi_\gamma = \Phi(x|\bmu_\gamma, \bSigma_\gamma)$
and the mixing weights $v_\gamma$ for $\gamma \in [K] $.
In vector format, the weights are $\bw$ and $\bv$.
{\it Be aware that by abuse of notation $\Phi_\gamma \neq \Phi_i$ even if $\gamma = i$.}

Let $c(\cdot, \cdot)$ be a cost function based on 
a divergence in the space of Gaussian distributions
(with a given dimension) for which the computational cost is low.
Then the transportation divergence~\citep{nguyen2013convergence} 
between mixing distributions $\bar G$ and $G \in \bbG_K$ with cost function $c$ becomes
\[
\mathcal{T}_{c}(\bar G, G) 
=
\inf \left \{ 
\sum_{i, \gamma}  {{\pi}}_{i \gamma} c( \Phi_i , \Phi_\gamma): \bpi \in \Pi( \bw, \bv) \right \}.
\]
The corresponding GMR estimator is
\be
\label{eq:gmr_estimator_transport}
\bar{G}^R 
= 
\arginf \big \{  \mathcal{T}_{c}(\bar G, G) : G \in \bbG_K \big \}.
\ee
Based on~\eqref{eq:gmr_estimator_transport}, it may appear that calculating our estimator involves two optimizations: 
computing $\mathcal{T}_{c}(\bar G, G)$ for each pair of $\bar G$ and $G$, and optimizing to find
$\arginf_G \mathcal{T}_{c}(\bar G, G)$.
This is not the case, as our algorithm will show. The optimization in $\mathcal{T}_{c}(\bar{G}, G)$ involves
searching for transportation plans ${\bpi}$ under two marginal constraints
specified by $\bw$ and $\bv$. 
While constraint $\bw$ is strict, $\mathbf{v}$ is a moving constraint.
Instead of searching for  ${\bpi}$ satisfying constraint $\bv$,
we move  $\bv$ to meet ${\bpi}$.
This makes the marginal distribution constraint
$\mathbf{v}$ on ${\bpi}$ redundant. 

Let us define two functions of $G$, with $\bar G$ hidden in the background:
\begin{align}
\label{Jcal}
\mathcal{J}_{c}(G) 
&=
\inf_{\bpi}  \big \{ 
\sum_{i, \gamma}  {\pi}_{i \gamma} c( \Phi_i , \Phi_\gamma): 
\bpi \in \Pi( \bw, \cdot)   \big \}, 
\\
{\bpi}(G) 
\label{bpi-star}
& = 
\arginf_{\bpi} \big \{ 
\sum_{i, \gamma}  {\pi}_{i \gamma} c( \Phi_i , \Phi_\gamma): 
\bpi \in \Pi( \bw, \cdot)    \big \}.
\end{align}
Note that both functions depend on $G$ through its subpopulations
$\Phi_\gamma$ but are free of its mixing weights $\bv$. 
The optimizations in~\eqref{Jcal} and~\eqref{bpi-star}
involve only the linear constraint in terms of $\bw$.
Hence, the optimal transportation plan $\bpi(G)$ for a given $G$ has an analytical form:
\be
\label{eq:bpi}
\pi_{i \gamma}(G) 
= 
\begin{cases}
w_i & \text{if } \gamma = \argmin_{\gamma'} c(\Phi_{i}, \Phi_{\gamma'})\\
0 & \text{otherwise}.
\end{cases}
\ee

If $\argmin_{\gamma'}c(\Phi_{i}, \Phi_{\gamma'})$ is not unique 
which is rare in practice, we may randomly pick one from the solution set.

\begin{theorem}
\label{thm:ww_averaging_equivalent_obj}
Let $\bar G$, $\mathcal{T}_{c}(\cdot)$, $\mathcal{J}_{c}(\cdot)$, $\bpi(\cdot)$,
and the other notation be as given earlier. 
We have
\begin{equation}
\label{eq:equiv_optimization}
\inf\{\mathcal{T}_{c}( G): G \in \mathbb{G}_{K}\} 
= 
\inf\{\mathcal{J}_{c}(G): G \in \mathbb{G}_{K}\}.
\end{equation}
The subpopulations of the GMR estimator are hence given by 
\begin{equation}
\label{eq:gmr_estimator}
\bar{G}^R 
= \arginf\{\mathcal{J}_{c}(G): 
  G \in \mathbb{G}_{K}
\}
\end{equation}
and the mixing weights are given by 
\begin{equation}
\label{eq:gmr_estimator_weight}
\mathbf{v} = \sum_{i} \bpi(\bar{G}^{R}).
\end{equation}
\end{theorem}

The existence of a solution to~\eqref{eq:gmr_estimator} is guaranteed 
under a simple condition on cost function $c(\cdot,\cdot)$, see Theorem~\ref{thm:convergence}.
The proof of Theorem~\ref{thm:ww_averaging_equivalent_obj} is in Appendix~\ref{sec:app_equivalent_obj}.
Based on this theorem, the optimization reduces to
searching for $K$ subpopulations $\Phi_\gamma$ for $\gamma \in [K]$
to make up $G$. 
The mixing proportions are then determined by~\eqref{eq:gmr_estimator_weight}.
An iterative algorithm quickly emerges following the
well-known majorization--minimization (MM) idea~\citep{hunter2004tutorial}.

The algorithm starts with some $G^{(0)}$ with $K$ subpopulations specified.
Let $G^{(t)}$ be the mixing distribution after $t$ MM iterations. 
Define a majorization function of $\mathcal{J}_{c}$ at $G^{(t)}$  to be
\be
\label{eq:majorization_function}
\mathcal{K}_{c}(G|G^{(t)}) 
= 
\sum_{i,\gamma} \pi_{i\gamma}(G^{(t)}) c(\Phi_{i}, \Phi_{\gamma})
\ee
where $\pi_{i\gamma}(G^{(t)})$ is computed according to~\eqref{eq:bpi}.
Once $\bpi(G^{(t)})$ has been obtained, we update the mixing proportion
vector of $G^{(t)}$ easily via
\begin{equation*}
\label{eq:weight_update}
v ^{(t+1)}_\gamma = \sum_i \pi_{i \gamma} (G^{(t)}).
\end{equation*}
In fact, $\bv^{(t)}$ is not needed until the algorithm concludes,
when it becomes an output.

The subpopulations $\Phi_{\gamma}$ are separated in the majorization 
function~\eqref{eq:majorization_function}.
This allows us to update the subpopulation parameters, one $\Phi_\gamma$
at a time and possibly in parallel,
as the solutions to 
\be
\label{eq:support_update}
    \Phi_{\gamma}^{(t+1)} 
    = \arginf_{\Phi} \sum_{i} \pi_{i\gamma}(G^{(t)})c(\Phi_{i}, \Phi).
\ee
The MM algorithm then iterates between the
majorization step~\eqref{eq:majorization_function}
and the
maximization step~\eqref{eq:support_update}
until some user-selected convergence criterion is met.

The most expensive step in the MM algorithm
is the optimization~\eqref{eq:support_update}. 
If we choose the cost function $c(\cdot, \cdot) = \rho^{r}(\cdot, \cdot)$ with
$\rho(\cdot, \cdot)$ being a divergence in the space of probability
measures, the solution to~\eqref{eq:support_update} is a barycenter as given in Definition~\ref{def:barycenter}.

\begin{lemma}[KL-Barycenter of Gaussian Measures]
\label{Gaussian-barycenter}
Let 
$\nu_m = \Phi(\cdot |\bmu_m, \bSigma_m), m \in [M]$ and 
$\blambda =(\lambda_1, \ldots, \lambda_M) \in \Delta_{M}$. 
Then 
$\sum_{m=1}^M \lambda_m D_{\text{KL}}(\nu_m\|\eta)$ 
is minimized uniquely in the space of Gaussian measures 
at $\eta = \Phi(\cdot |\bar{\bmu} , \bar{\bSigma})$
with
\begin{equation*}
\bar{\bmu}= \sum_{m=1}^M\lambda_m \bmu_m
~~\mbox{ and }~~
\bar{\boldsymbol\Sigma} 
= 
\sum_{m=1}^M \lambda_m 
\{
\bSigma_m + (\bmu_m-\bar{\bmu})(\bmu_m-\bar{\bmu})^\tau
\}.
\end{equation*}
\end{lemma}

The above lemma shows that the KL-based barycenter has an analytical
form and is therefore computationally simple. 
Therefore, we choose $c(\Phi_{i},\Phi_{\gamma}) = \dKL(\Phi_{i}\|\Phi_{\gamma})$ 
in~\eqref{eq:gmr_estimator_transport} for its simplicity. 
The pseudo-code for this cost function is given in 
Algorithm~\ref{alg:mm_reduction} in Appendix~\ref{app:algorithm}.
Different cost functions are associated with different geometries 
on the Gaussian distribution space~\citep{peyre2019computational}, 
and other cost functions can be used if desired. 
\todo{See~\citet{zhang2020unified} for a discussion of different cost functions.}

For notational convenience, in the following theorem, we use
$\Phi$ for both the parameter $(\bmu, \bSigma)$ and the distribution,
and similarly for $\Phi^*$.

\begin{theorem} 
\label{thm:convergence}
Suppose the cost function $c(\cdot, \cdot)$ is continuous in both arguments.
For some distance in the parameter space of $\Phi$, assume that for any constant $\Delta > 0$ and $\Phi^*$ the following set is compact:
\be
\label{eq:compact.Phi}
\{ \Phi: c(\Phi^*, \Phi) \leq \Delta \}
\ee
Let $\{G^{(t)}\}$ be the sequence generated by  
$G^{(t+1)} = \argmin \mathcal{K}_{c}(G|G^{(t)})$ 
with some initial mixing distribution $G^{(0)}$.
Then
\begin{itemize}
    \item [(i)] 
    $\mathcal{J}_c(G^{(t+1)}) \leq \mathcal{J}_c(G^{(t)})$ for any $t$;
    \item[(ii)]
    if $G^*$ is a limiting point of $G^{(t)}$, then $G^{(t)} = G^*$ implies 
    ${\cal J}_c(G^{(t+1)}) = {\cal J}_c(G^*)$.
    \end{itemize}
\end{theorem}
\noindent
These two properties imply that $\mathcal{J}_{c}(G^{(t)})$ converges monotonically 
to some constant $\mathcal{J}^*$. 
All the limiting points $G^{(t)}$ are stationary points of $\mathcal{J}_c(\cdot)$:
iterations from $G^*$ do not further reduce the value of the objective function $\mathcal{J}_c(\cdot)$.
We have practically cloned
the Global convergence theorem~\citep{zangwill1969nonlinear}
here, but we do not see a way to directly apply it.
The proof of the theorem is given in 
Appendix~\ref{app:condition_for_global_convergence_theorem}.

In principle, we now have all the ingredients for the split-and-conquer
method for the learning of a finite GMM.
We next consider the statistical properties of
the GMR estimator and the experimental evidence for the
efficiency of our method.

\section{Statistical Properties of the Reduction Estimator}
\label{sec:theory}

We now show that the proposed GMR estimator $\bar{G}^{R}$ is consistent
and retains the optimal rate of convergence in a statistical sense.
We first state some conditions on the data and the estimation methods.
\begin{itemize}
\label{page.C}
\item[\bf C1]
The data $\Xcal$ are IID observations from the finite Gaussian mixture distribution $\Phi(x; G^*)$ with
$K$ distinct subpopulations, that is the order of $G^*$ is known to be $K$. 
The subpopulations have distinct parameters and positive definite covariance matrices.
\item[\bf C2]
The  dataset $\Xcal$ is partitioned into $M$ subsets $\Xcal_1, \ldots, \Xcal_M$
of sizes $N_1, \ldots, N_M$, where each set contains IID observations
from the same finite Gaussian mixture distribution. 
The number of local machines $M$ does not increase with $N = \sum_m N_m$.
\item[\bf C3]
The local machine sample ratios
$N_m/N$ have a nonzero limit as $N\rightarrow\infty$.
\item [\bf C4] 
The cost function $c(\Phi_k, \Phi_0) \to 0$ or $c(\Phi_0, \Phi_k) \to 0$ 
if and only if $\Phi_k \to \Phi_0$ in distribution, and
 $c(\Phi_1, \Phi_2)$ is continuous in both $\Phi_1$ and $\Phi_2$.
\end{itemize}

Condition C4 is necessary to ensure consistency. It further rules out the case that
$\mathcal{T}_{c}(G, G^*) = \infty$ for any $G$ with a
different set of mixing weights from that of $G^*$.

The consistency and asymptotic normality of the pMLE under the finite multivariate 
GMM are established in~\citet{chen2009inference}
under the standard IID condition. We do not repeat
the details here but state the conclusion in a simplified fashion.
We use $\Phi_k^*$, $(w^*_k, \bmu^*_k, \bSigma^*_k) $ to denote
true subpopulation, the mixing weight and the subpopulation parameters.

\begin{lemma}[Consistency of pMLE]
\label{lemma:consistency_pMLE}
Given $n$ IID observations from a finite multivariate GMM
with known order $K$, the pMLE $\hat G$ as defined by \eqref{eq:pMLE}
with $a_n = n^{-1/2}$ is asymptotically normal with rate $n^{-1/2}$.

Specifically, it is possible to line up the subpopulation
parameters of   the true mixing distribution $G^*$
and of the pMLE $\hat G$ such that
\[
(\hat w_k, \hat \bmu_k, \hat \bSigma_k) =
(w^*_k, \bmu^*_k, \bSigma^*_k) + o(1)
\]
and
\[
(\hat w_k, \hat \bmu_k, \hat \bSigma_k) =
(w^*_k, \bmu^*_k, \bSigma^*_k) + O_p(n^{-1/2})
\]
as $n \to \infty$ in obvious notation.
\end{lemma}

Here $o(1)$ is a quantity that converges to $0$ almost surely.
A quantity is $O_p(n^{-1/2})$ if it is bounded by $Cn^{-1/2}$ for
sufficiently large $C$ with a probability arbitrarily close to 1.~\citet{chen2009inference} prove the root-n-consistency for a non-random penalty term, the asymptotic normality remains valid when the sample covariance matrix is part of the penalty.
This is because the sample covariance matrix converges to a positive definite matrix. 
The assumption of known $K$ is crucial for the claimed
rate of convergence. If $K$ is unknown, then the
convergence rate of $\hat G$ is far below $n^{-1/2}$: see~\citet{chen1995optimal}
and the recent developments in \citet{rousseau2011asymptotic}, \citet{nguyen2013convergence}, \citet{heinrich2018strong}, 
and \citet{dwivedi2020sharp}.
Our proposed reduction estimator is aggregated from the pMLEs learned at
the local machines. 
Under the condition that $\min N_m \to \infty$ stated above, all the local
estimators are consistent by Lemma \ref{lemma:consistency_pMLE}. 
Hence, the consistency of the aggregate estimator $\bar G$ is taken as granted for a constant $M$.

\begin{theorem}[Consistency of $\bar{G}^{\text{R}}$] 
\label{GMR-consistency-KL}
Let $\bar G$ be the aggregate estimator defined by \eqref{eq:pool_estimator}
and $\bar{G}^{R}$ be the aggregate estimator by reduction defined by~\eqref{eq:reduction}
with $\rho = \mathcal{T}_{c}$.
Assume Conditions C1--C4 are satisfied. 
Then $\bar G^R$ is strongly consistent. Specifically,
\(
\Tcal_{c}(\bar G^R, G^*)  \to 0
\)
almost surely as $N \to \infty$.
\end{theorem}

The proof of the theorem is given in Appendix~\ref{app:consistency_proof}. 
The following theorem shows that under one additional 
mild condition on the cost function $c(\cdot,\cdot)$, 
the reduction estimator $\bar G^R$ has the standard $N^{-1/2}$ convergence rate.
We denote by $\| \Phi_1 - \Phi_2\|$ the Euclidean norm in $\bmu, \bSigma$
in the sense of \eqref{D.behind.W1}. We use $\bar \Phi_k^R$ for the $k$th subpopulation of $\bar G^R$ and $\bar w_k$ for its mixing weight, $k \in [K]$.

\begin{theorem}[Convergence rate of $\bar{G}^{\text{R}}$]
\label{thm:rate_of_convergence}
Let $\bar G$ be the aggregate estimator defined by \eqref{eq:pool_estimator}
and $\bar{G}^{R}$ be the aggregate estimator by reduction defined by~\eqref{eq:reduction}.
Assume Conditions C1--C4 are satisfied and further assume that
\begin{itemize}
\item [\bf C5]
For any $\Phi$, there exists a small neighborhood $\Omega$ of $\Phi$ 
and a positive constant $A$,
such that for any $\Phi_1, \Phi_2 \in \Omega$, we have
\[ 
A^{-1} \| \Phi_1 - \Phi_2\|^2 \leq c(\Phi_1, \Phi_2)  \leq  A \| \Phi_1 - \Phi_2\|^2.
\]
\end{itemize}
Then with proper labelling of subpopulations, we have
\[
\bar \Phi_k^R - \Phi^*_k = O_p(N^{-1/2}), ~~
\bar w_k^R - \pi^*_k = O_p(N^{-1/2}).
\]
\end{theorem}

Condition C5 requires the cost function $c(\cdot, \cdot)$ to
behave locally as a quadratic loss function. This is a most natural
property a cost function should have. In Appendix~\ref{app:kl_divergence_property}, we show this condition holds for the KL divergence. 
The conclusion should hold with any other reasonable choices. 
The proof of the theorem is given in Appendix~\ref{app:convergence_rate_proof}. 
Our proof remains valid, for instance, if we replace $\|\Phi_1 - \Phi_2\|^2$
by $\|\Phi_1 - \Phi_2\|^r$ for any $r > 0$ in C5.

\section{Experiments}
\label{sec:experiment}
In this section, we use experiments with both simulated and real data to 
illustrate the effectiveness of our GMR estimator in~\eqref{eq:gmr_estimator_transport} with $c(\Phi_i, \Phi_\gamma) = \dKL(\Phi_i, \Phi_\gamma)$. We compare its performance with existing approaches
in terms of their statistical efficiency and computational costs. 
We consider the following estimators:
\begin{enumerate}
\item \emph{Global}.
The pMLE based on the full dataset. This is conceptually the most efficient, and
its performance gives us a gold standard.

\item \emph{Median}. We follow the principle of split-and-conquer principle and use the median estimator defined below for aggregation, that is
\begin{equation*}
\label{eq:median_estimator}
\bar{G}^{\text{M}} 
= \argmin_{G \in \{\hat{G}_1,\hat{G}_2,\ldots, \hat{G}_M\}} 
\sum_{m=1}^{M} \lambda_m \mathcal{T}_{\dKL} (\hat{G}_m, G). 
\end{equation*}
The sample median is robust to outliers at the cost of possible efficiency loss. 

\item 
\emph{KLA}. We consider to compare with the KLA in~\citet{liu2014distributed}. In the simulated data experiment, we generate $n_m=1000$ observations from $\hat G_m$ learned on the $m$th local machine. Since the real datasets have different dimensions and sample sizes, the size of the generated samples in each real data experiment is specified when the real dataset is described.

\item 
\emph{Coreset}. For the coreset estimator, in the simulated data experiment, we construct a coreset $\gC_{m}$ with $|\gC_{m}|=1000$ on each local machine and merge these coresets in pairs of two to a coreset with size $1000$ as in~\citet{lucic2017training}. This leads to a merged coreset $\gC$ with size $1000$. Same reason as the KLA estimator, the coreset size on the real data is specified in the corresponding section.
\end{enumerate}

For the ease of comparison, we use the same local estimators for all split-and-conquer based estimators.
We use the pMLE defined in \eqref{eq:pMLE} at each local machine 
with the size of the penalty set to $N_m^{-1/2}$.
When computing the KLA estimator of~\citet{liu2014distributed}
on the central machine, we choose the penalty size by the same rule,
that is $(1000M)^{-1/2}$.
We also fit the mixture based on the coreset on the central machine with pMLE, the penalty size is chosen to be $|\gC|^{-1/2}$. 
We use the EM-algorithm to compute pMLE.
We declare convergence when the increment in the penalized log-likelihood function 
standardized by the total sample size is less than $10^{-6}$.
Since the sample sizes in our simulated data experiment are very large, the maxima of the penalized likelihood should be attained at a mixing distribution close to the actual mixing distribution.  
In the simulation experiments, we can therefore use the true mixing distribution as the initial value and regard the output of the EM-algorithm as the global maximum of the penalized likelihood. 
This strategy does not work for the real data in the absence of a true mixing distribution.
We run the EM algorithm as follows, we use \emph{kmeans++} to generate 10 initial values for the EM-algorithm using the \emph{scikit-learn} package~\citep{scikit-learn}.
Ideally, we run the EM algorithm with these initial values until converge and regard the output of the EM algorithm with the highest penalized log-likelihood function as the pMLE.
However, to run the EM algorithm for many rounds on large datasets can be very time consuming.
To save the computational time on the real dataset, we use the warmup strategy. 
We run the EM algorithm with these 10 initial values for $20$ iterations and pick the output of the one with the highest penalized log-likelihood value as the initial value to run the EM algorithm until converge and the output is treated as the pMLE.

The choice of the initial value in the GMR approach is also important 
since the objective function is non-convex.
When the sample size is large, we have good reason to believe that
the optimal solution is close to the true value. Also, by the principle of majority rule, the median of the local estimates is likely the closest to the optimal solution.
Thus, in the simulation studies, we initialize the algorithm with the true mixing distributions or the median estimator.
For real data, we use the local estimators as the initial values and output of the MM algorithm with the lowest objective function value as the GMR estimator.
We declare the convergence of the MM algorithm for the GMR estimator 
when the change in the objective function is less than $10^{-6}$.

All the experiments are conducted on the Compute Canada~\citep{baldwin2012compute} Cedar cluster with Intel E5 Broadwell CPUs with 64G memory. 
All experiments are implemented in Python and the code is publicly available at \url{https://github.com/SarahQiong/SCGMM}.
The code for the coreset method is provided by the authors of~\citet{lucic2017training}.


\subsection{Performance Measure}
In the simulation, we generate $R = 100$ random samples $\Xcal^{(r)}$ from a finite GMM with mixing distribution $G^{(r)}$ for $r \in [R]$.
The following two loss functions are used to measure the performance of the various estimation methods. 
\begin{enumerate}
\item 
\textbf{Transportation divergence} ($W_1$).
For any two Gaussian distributions $\Phi_{i}$ and $\Phi_{\gamma}$
with mean vectors $\bmu_{i}$ and $\bmu_{\gamma}$ and covariance matrices $\bSigma_i$ and $\bSigma_{\gamma}$, we define a distance
\be
\label{D.behind.W1}
D(\Phi_{i}, \Phi_{\gamma}) 
=
 \|\bmu_{i}-\bmu_{\gamma}\|_{2} + \|\bSigma_i^{1/2}-\bSigma_{\gamma}^{1/2}\|_{F}
\ee
where $\|A\|_F = \sqrt{\text{tr}(A^{\tau}A)}$ is the Frobenius norm of a matrix.
The corresponding transportation distance is
$$
\Lcal(\hat{G}^{(r)}, G^*)= 
\inf \left\{ \sum_{i,\gamma}\pi_{i \gamma} D( \Phi_i, \Phi_\gamma) : \bpi\in\Pi(\mathbf{w},\mathbf{v}) \right\}$$
where $\Phi_i$ and $\Phi_\gamma$ are subpopulations
in $\hat{G}^{(r)}$ and $G^{(r)}$ respectively. The mixing weights of $\hat{G}^{(r)}$ and $G^{(r)}$ are $\mathbf{w}$ and $\mathbf{v}$ respectively.

\item \textbf{Adjusted Rand Index} (ARI). 
The mixture models are usually used for the purpose of clustering based on the rule of maximum posterior.
That is
\begin{equation}
\label{eq:clustering}
\hat k^{(r)}(\vx)
= \argmax_{k} \{\hat w_k \phi(\vx | \hat{\bmu}_k, \hat{\bSigma}_k)\}
\end{equation}
based on the learned mixing distribution $\hat G^{(r)}$.
Similarly, the clustering outcome $k^{(r)}(\vx)$ based on the true mixing distribution $G^{(r)}$ for each observation can be obtained. 
We measure the performance of the learning approach by measuring the degree of similarity between $\{\hat{k}^{(r)}(\vx_i):i=1,2,\ldots,N\}$ and $\{k^{(r)}(\vx_i):i=1,2,\ldots,N\}$ by the adjusted Rand index (ARI).
Suppose the observations in a dataset are divided into $K$ clusters
$A_1, A_2, \ldots, A_K$ by one method,
and $K'$ clusters $B_1, B_2, \ldots, B_{K'}$ by another method. 
Let $N_i = \# (A_i), ~ M_j = \# (B_j), ~~ N_{ij} = \# (A_i B_j)$ for $i \in [K]$ and $j \in [K']$, where $\#(A)$ is the number of units in set $A$.
The ARI of these two clustering outcome is defined to be
\[
\text{ARI}= \dfrac{
\sum_{i, j } \binom{N_{ij}}{2}
-  \binom{N}{2}^{-1} \sum_{i, j}\binom{N_{i}}{2}\binom{M_{j}}{2}}
{\frac{1}{2} \sum_{i}\binom{N_{i}}{2} + \frac{1}{2}  \sum_{j} \binom{M_{j}}{2}
- \binom{N}{2}^{-1} \sum_{i, j}\binom{N_{i}}{2}\binom{M_{j}}{2}}.
\]
When the two clustering methods completely agree with each other,
\text{ARI} takes the value 1. Values close to 1 indicate
a high degree of agreement. 

\item \textbf{Log-likelihood} (LL). We also evaluate the log-likelihood function at the estimated mixing distribution $\hat{G}$ based on the full generated dataset. For the ease of presentation, we present the value of the log-likelihood per observation.
\end{enumerate}

The split-and-conquer approach may also reduce the computational time by performing local inference on multiple machines. We therefore also report the computational times of all the methods.
The computational times of the split-and-conquer approaches
are defined to be the sum of the time for the local estimates 
and that for the aggregated estimator. Since the local estimates can be computed in parallel, 
we record the longest local machine time as the time for the local estimates.
The computational time for the coreset is the time for constructing the coreset and the time for fitting the mixture on the central machine, we ignore the time for transmitting the dataset from local machine to central machine.

\subsection{Experiments Based on Simulated Data}
\label{sec:simulated_data}
\begin{figure}[!htp]
\centering
\subfloat[Transportation distance $W_1$]{\includegraphics[width = 0.95\textwidth]{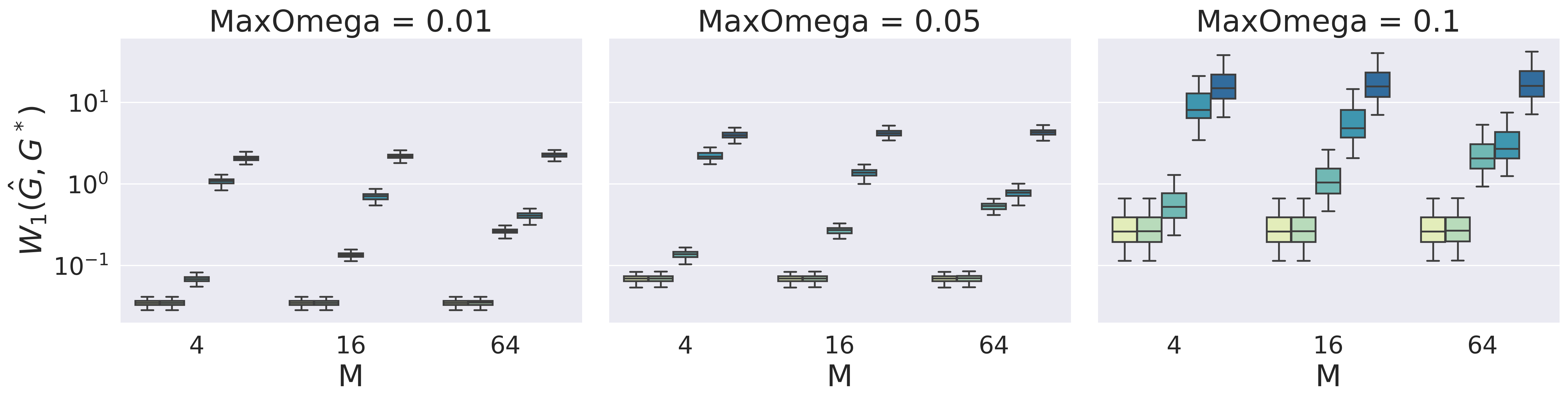}}\\
\subfloat[ARI]{\includegraphics[width = 0.95\textwidth]{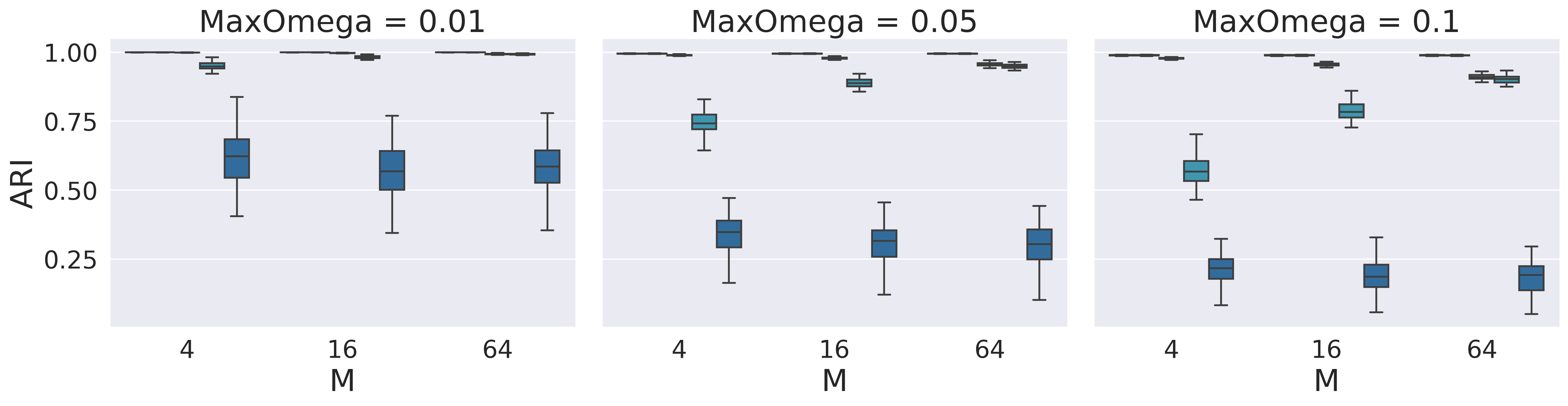}}\\
\subfloat[Log-likelihood]{\includegraphics[width = 0.95\textwidth]{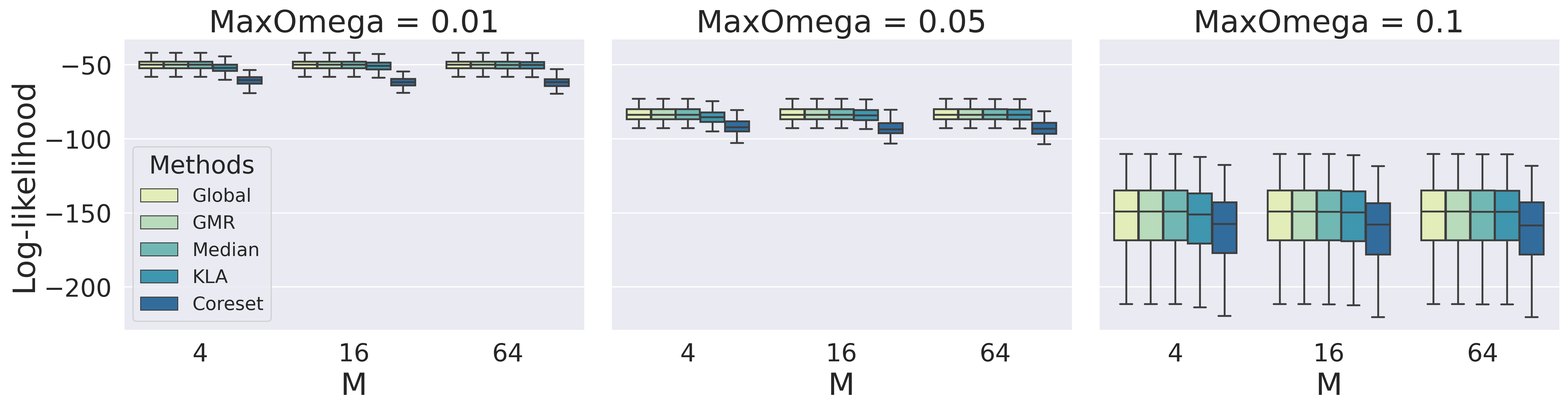}}\\
\subfloat[Computational time]{\includegraphics[width = 0.95\textwidth]{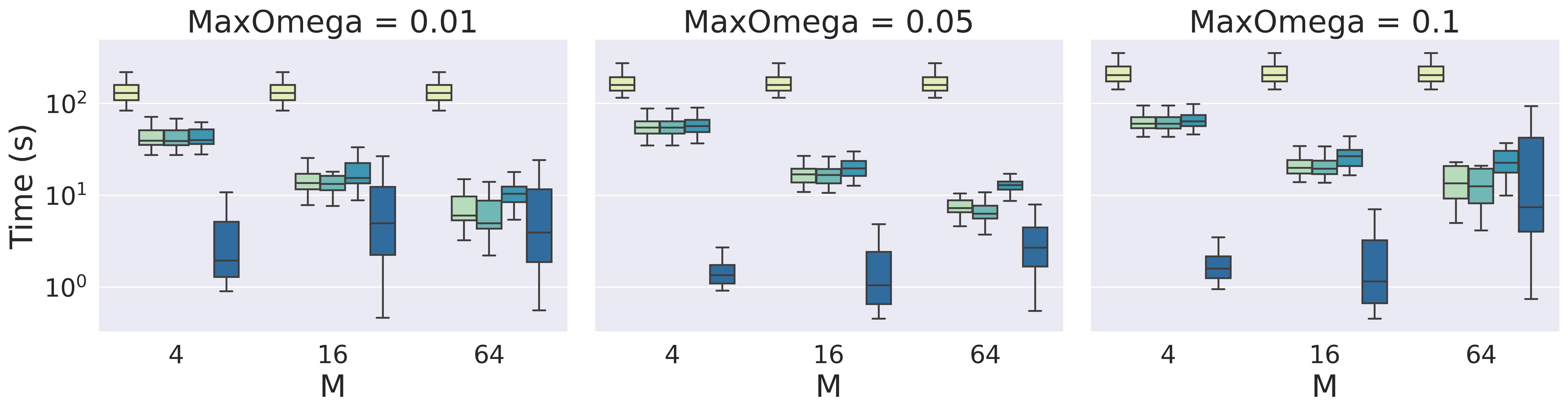}}
\caption{Comparison of Global, Median, GMR, KLA, and Coreset estimators 
for learning of $50$-dimensional $5$-component mixture when sample size is $N=2^{21}$ for various number of local machines and degree of overlap.}
\label{fig:local_machine_ss}
\end{figure}

\begin{figure}[!htp]
\centering
\subfloat[$W_1$ distance]{\includegraphics[height = 0.25\textwidth]{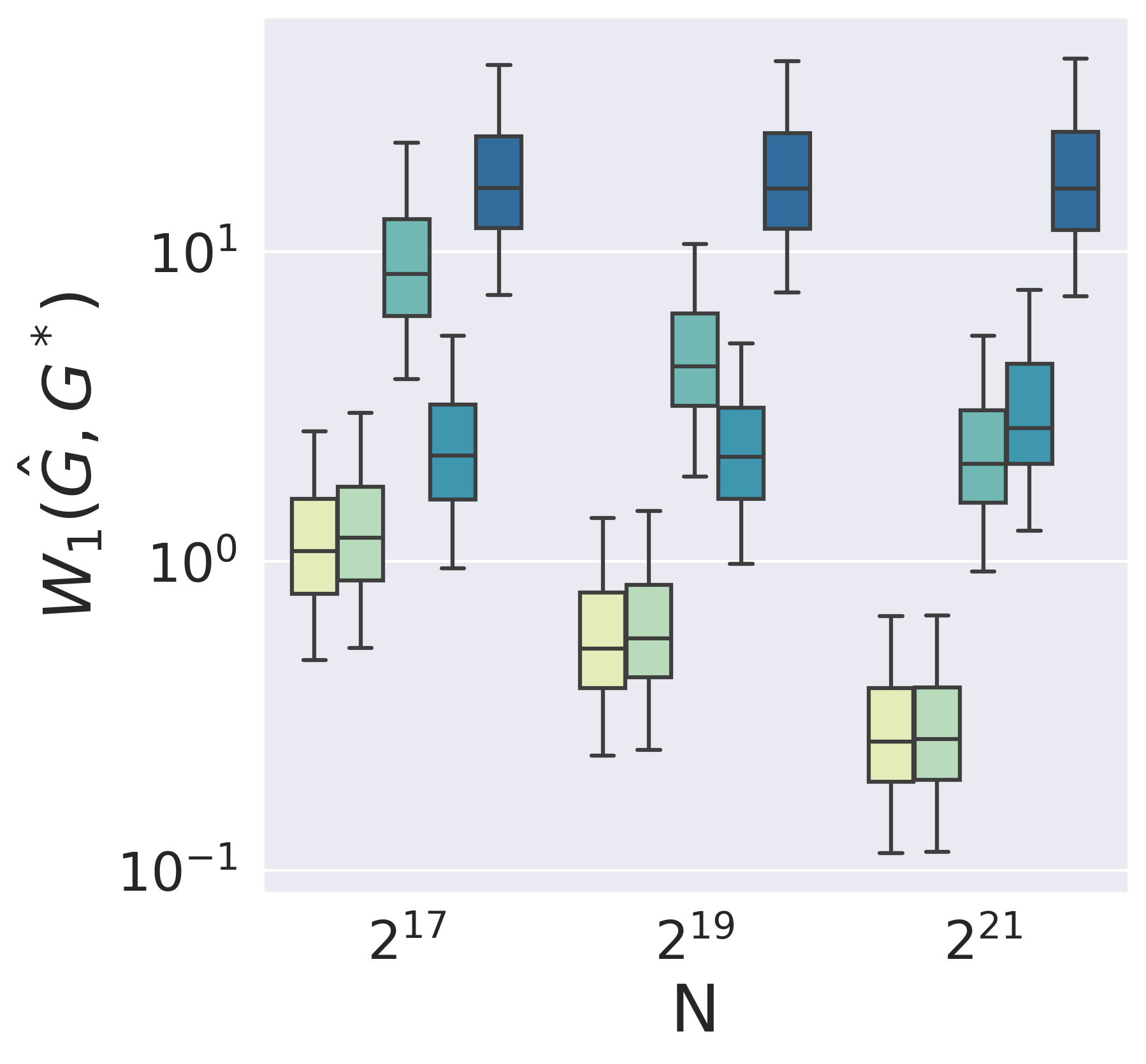}}
\subfloat[ARI]{\includegraphics[height = 0.25\textwidth]{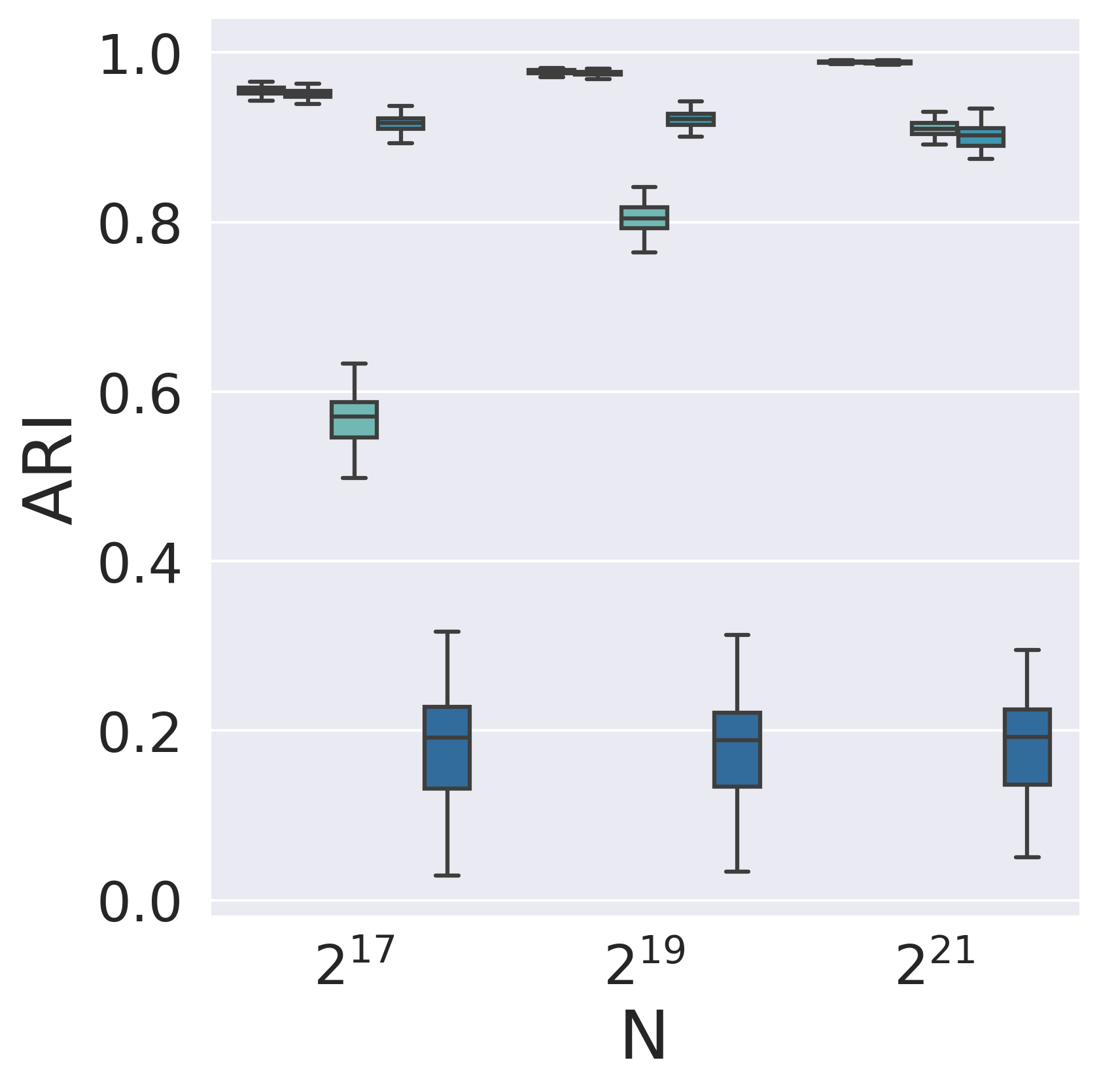}}
\subfloat[Log-likelihood]{\includegraphics[height = 0.25\textwidth]{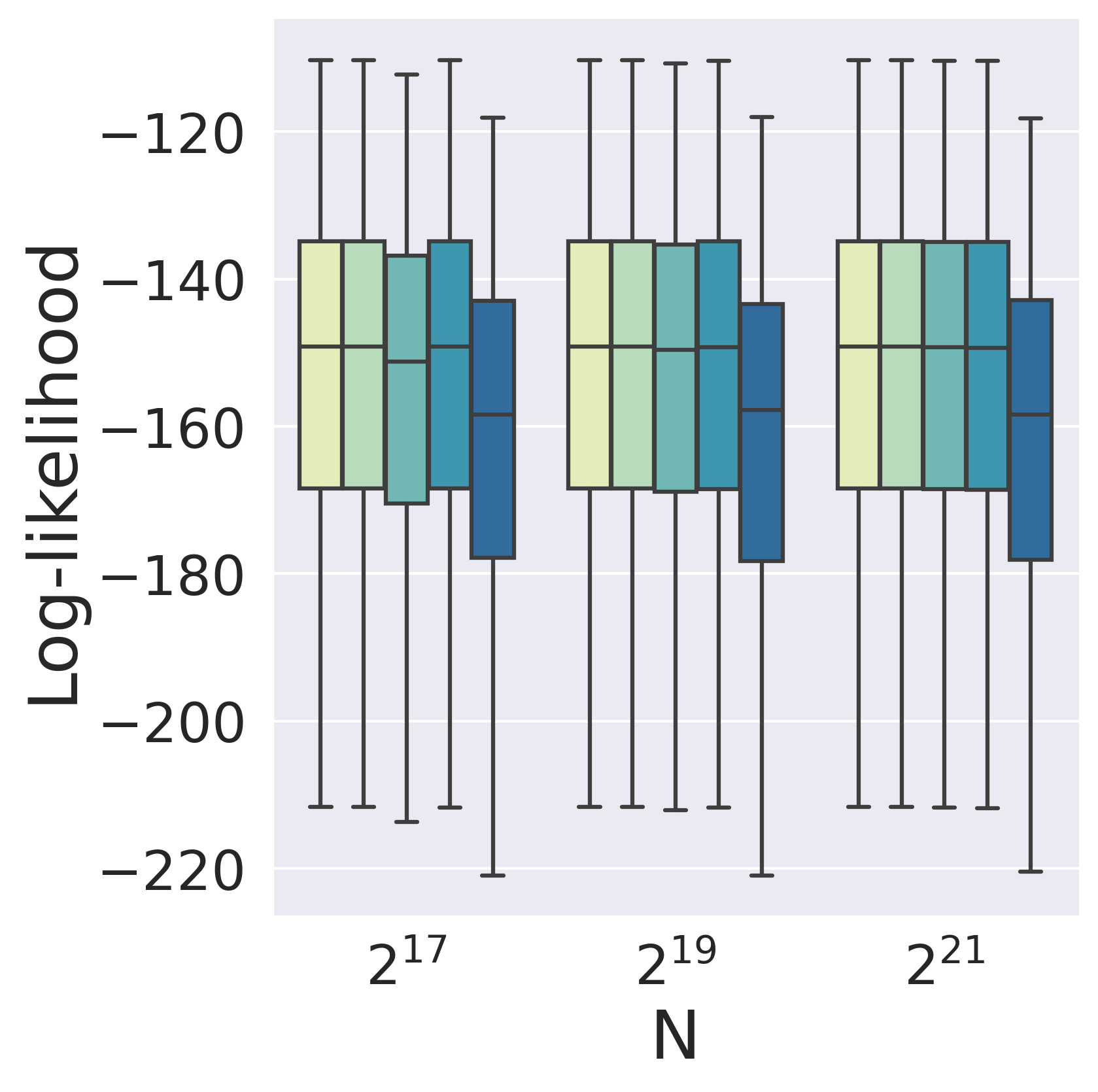}}
\subfloat[Computational time]{\includegraphics[height = 0.25\textwidth]{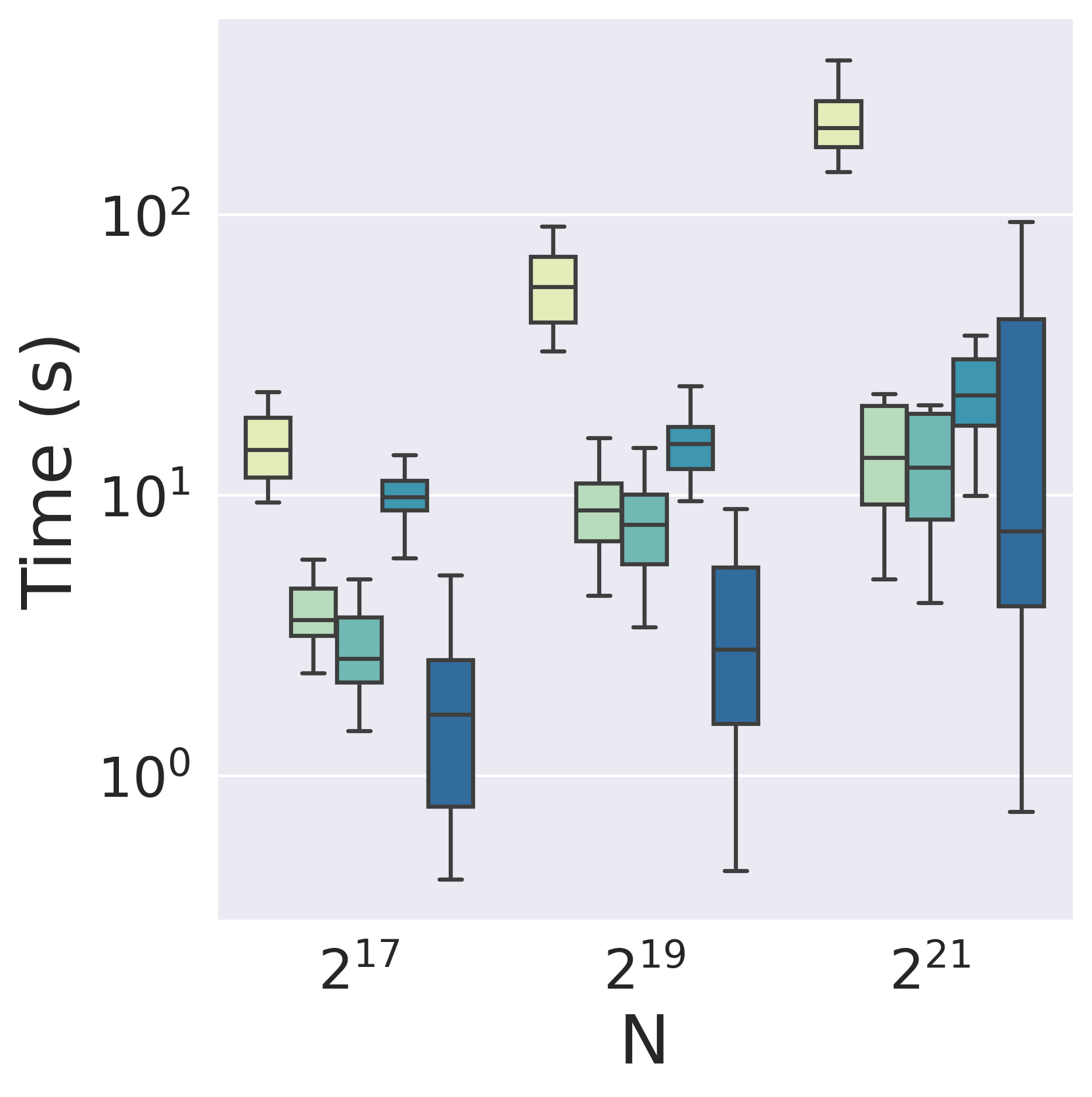}}
\caption{Comparison of Global, Median, GMR, KLA, and Coreset estimators when \texttt{MaxOmega} $=0.1$ and $M=64$ for learning of 
$50$-dimensional 
$5$-component mixture for various sample sizes.}
\label{fig:ss_estimator_comparison}
\end{figure}

Distributed learning methods for finite mixtures are intended for large
problems where the observations
have high dimensions. Our experiments are biased toward such situations.
To reduce the potential influence of human bias, 
we simulate data from finite GMMs for which the parameters
themselves are randomly generated. 
We use the \texttt{R} package \texttt{MixSim}~\citep{melnykov2012mixsim}, 
which is based on~\citet{maitra2010simulating}.
An important feature of a finite mixture model is
the maximum degree of pairwise overlap. 
Let 
\[
o_{j|i} 
= \mathbb{P}(w_{i}\phi(X|\bmu_{i},\bSigma_{i}) < w_{j}\phi(X|\bmu_{j},\bSigma_{j}) 
 |X\sim N(\bmu_{i},\bSigma_{i})),
\]
which is the probability of a unit from subpopulation $i$ being misclassified into
subpopulation $j$ by the model-based classification rule.
The degree of overlap between the $i$th and $j$th subpopulations 
is therefore $o_{ij} = o_{j|i} + o_{i|j}$, and the maximum overlap is given by
\[
\texttt{MaxOmega} = \max_{i, j \in [K] } \{ o_{j|i} + o_{i|j} \}.
\]
We used \texttt{MixSim} to generate $100$ finite GMM distributions
of dimension $d=50$ with $K=5$ subpopulations and 
\texttt{MaxOmega} set to $1\%$, $5\%$, and $10\%$.
In this experiment, we set the sample sizes to  
$N=2^l$ for $l=17, 19, 21$ and the number of local machines to
$M=2^l$ for $l = 2, 4, 6$. The simulated data are distributed
evenly over the local machines. We use all five estimation methods: 
Global, GMR, Median, KLA, and Coreset.  
We combine 100 outcomes from each combination of dimension,
order, \texttt{MaxOmega}, sample size, and number of local machines
to form boxplots for each estimation method.
Figures~\ref{fig:local_machine_ss} and \ref{fig:ss_estimator_comparison} show the results.

In Figure~\ref{fig:local_machine_ss}, the total sample
size is fixed to $N=2^{21}$.
Within each subfigure, the \texttt{MaxOmega} level increases from the left panel to the right panel.
Within each panel, the x-axis gives the number of local machines: $4$, $16$, or $64$. Figures~\ref{fig:local_machine_ss}(a) to \ref{fig:local_machine_ss}(d) show the results for $W_1$, the ARI, the log-likelihood per observation, and the computational time respectively.
Due to the bad performance of the Coreset estimator compared to the rest of the estimators in terms of $W_1$ and ARI, the difference of the rest of the four methods are difficult to tell, we insert a small figure to only compare these four methods when necessary, for example the comparison of $W_1$ in the left subfigure in Figure~\ref{fig:local_machine_ss}(a).

In general, the $W_1$ distances increase, the ARIs and log-likelihood decrease as the degree of overlapping increases. 
Our GMR estimator has comparable performance with the gold-standard global estimator in terms of both $W_1$ distance, ARI, and log-likelihood. 
The number of local machines has little influence on its performance. 
The KLA has even worse performance than the median, but it does better as the number of local machines increases.
This is because the total sample size at the
central machine increases proportionally with the number of local machines.
The median estimator does not perform well when the number of local machines increases, likely because of the decreased sample sizes on the local machines.
The Coreset estimator is worse than the KLA and its performance gets worse as the number of local machines increases, this is likely due to the information loss occurred during the merge of the coresets. The information loss increases as the number of merges increases.

In terms of computational time, all the aggregation approaches are more efficient than the global estimator and the Coreset estimator is computational fastest.
However, the computational time for the Coreset estimator ignores the time for transmission, the actual computational time is therefore longer.
The times increase as the degree of overlapping increases since more iterations are often needed. 
They also increase as the number of local machines increases.
Because the aggregation times for GMR and the median estimators are often negligible, they use around (1/M)th of the time required by the global estimator. 
The aggregation step is more time-consuming for KLA, but it is still low.
It has to fit the GMM an additional time on the pooled data generated
from the local estimates.

In Figure~\ref{fig:ss_estimator_comparison},
\texttt{MaxOmega} is fixed to $10\%$ and the number of local machines is set to $M=64$ for the same dataset.
In general, Global, GMR, and Median
perform better as the total sample size increases. 
The performance of Coreset gets worse as total sample size increases, this is because the coreset size is fixed to be $1000$ in all experiment and the precision of the approximation decreases as the total sample size increases. 
KLA performs slightly worse as the total sample size increases.
Because the number of local machines is fixed, the larger sample size
improves only the precision of the local estimates. 
For KLA the aggregation step is based on the observations generated from each local estimate; their built-in randomness is an important factor.
The increased overall sample size is not particularly helpful.
The other aspects of the simulation results are as expected.

In Appendix~\ref{app:more_simulation_results} we present simulation results for dimension $d=10, 50$ combined with the order $K=5$, $10$, and $50$.
Since the performance of the Coreset estimator is not as good as the rest of the estimators in the simulation described above, we therefore do not compare it with the rest of the estimators in this experiment.
The general conclusions are similar in terms of the statistical efficiency. The local machine may take longer when $d=50$, $K=50$, and \texttt{MaxOmega} $\geq 5\%$.

\subsection{Real Dataset}
We compare the performance of the five estimators on large scale public datasets in Section~\ref{sec:public_dataset}. 
In Section~\ref{sec:NIST_dataset}, we show the performance of these five estimators on clustering the handwritten digits.
In Section~\ref{sec:CAM_dataset}, we show an example of using the distributed learning for the application of clustering in large scale spatio-temporal data.

\subsubsection{Public Datasets}
\label{sec:public_dataset}
We conduct the experiments on the public datasets that are widely used for learning Gaussian mixtures at scale.
The following datasets are used in~\citet{lucic2017training} and~\citet{jaini2016online}.
\begin{enumerate}
\item {\sc MAGIC04}. This is a simulated dataset to classify gamma particles in the upper atmosphere, which contains $19,020$ observations in a $10$-dimensional space. We fit a mixture of order $K=10$ on this dataset. The dataset is publicly available at UCI machine learning repository. 

\item {\sc MiniBooNE}. The dataset is taken from the MiniBooNE experiment that is used to distinguish electron neutrinos from muon neutrinos. The dataset is publicly available at UCI machine learning repository. The dataset contains $130,065$ observations in a $50$-dimensional space. We fit a mixture of order $K=10$ on the dataset. 

\item {\sc KDD}. This dataset is used in~\cite{lucic2017training} which contains $145751$ observations in a $74$-dimensional space for predicting the protein types. We fit a mixture of order $K=10$.
The KDD dataset is available at~\url{https://kdd.org/kdd-cup/view/kdd-cup-2004/Data}.

\item {\sc MSYP}. The dataset is used to predict the release year of a song from audio features. There are $515,345$ instances in $90$-dimensional space. Following~\citet{lucic2017training}, we consider the top $25$ principal components and fit a mixture with order $K=50$. 
The dataset is publicly available at UCI machine learning repository. 
\end{enumerate}
For the first three datasets, we split the dataset onto $M=4$ local machines completely at random. 
Since the last datasets have the largest number of observations and the order of the mixture is high, we split the dataset onto $M=16$ local machines.
The random split of the dataset is repeated for $R=100$ times for each split-and-conquer learning method.
The size of the generated sample and coreset size are set to $1000$ in MAGIC04, for the rest of the datasets, this value is set to $10000$.

We report the median and the IQR of the value of the log-likelihood per observation evaluated at the learned mixing distribution based on the full dataset over $100$ repetitions in Table~\ref{tab:real_data_ll}.
When evaluating the log-likelihood function on MiniBooNE dataset with KLA and Coreset estimator, there is a numerical underflow problem due to the extreme small size of the log-likelihood value, we therefore remove these observations and compute the log-likelihood per value based on the rest of the observations.
As a result, the upper bound of the log-likelihood value is given in the table.
The relative performance of these estimators on the dataset is the same as that on the simulated dataset.
It can be seen from the table that our proposed GMR approach is almost as good as the global estimator and is better than other split-and-conquer learning approaches on all datasets.
\begin{table}[htbp]
\centering
\tiny
\caption{The median log-likelihood per observation with IQR in parentheses.}
\begin{tabular}{cccccccccc}
\toprule
Dataset & $N$ & $d$ & $K$ & $M$ &Global& GMR& Median & KLA & Coreset\\
\midrule
MIGIC04 & 19020 & 10 & 10 &4& -24.15 &-24.30(0.07)&-26.60(0.05)&-26.73(0.07)&-27.16(0.55)\\
MiniBooNE & 130065 & 50 & 10 &4& -19.46& -22.00(0.53) & -24.60(0.32) & $\leq$-21.96(0.89)& $\leq$-25.11(3.21)\\
KDD &  145751&  74 & 10 &4&-221.80&-223.25(0.42)&-232.93(8.02)&-235.00(8.96)&-374.43(193.58)\\
MSYP & 515345 &25&50&16&-166.56&-167.05(0.04)&-171.10(0.04)&-170.72(0.01)&-181.64(1.78)\\
\bottomrule
\end{tabular}
\label{tab:real_data_ll}
\end{table}

The computational time of each method is given in Table~\ref{tab:real_data_time}, all the split-and-conquer learning methods can save a significant amount of time.
For the MSYP dataset, the split-and-conquer methods save as least 10 times time than the global estimator. The Coreset estimator takes the shortest time.
When the dimension of the mixture and the components of the mixture increases, the aggregation time of our proposed approach also increases but is still shorter than the KLA method. 
\begin{table}[htbp]
\centering
\tiny
\caption{The median computational time in seconds with IQR in parentheses.}
\begin{tabular}{cccccccccc}
\toprule
Dataset & $N$ & $d$ & $K$ & $M$&Global& GMR& Median & KLA & Coreset\\
\midrule
MIGIC04 & 19020 &  10 & 10 & 4&19.3& 7.0(3.2)&6.7(3.2)&10.2(3.1)&2.2(0.6)\\
MiniBooNE & 130065 &  50 & 10 & 4&346.9&313.1(162.6)&313.2(162.6)&511.3(213.2)&26.6(64.3)\\
KDD &  145751&  74 & 10 & 4&1033.9&544.4(309.5)&543.0(310.0)&706.0(290.3)&4.3(64.0)\\
MSYP & 515345 &25&50& 16&67048.8&2611.6(474.0)&1777.5(511.2)&5515.9(1629.7)&67.4(12.6)\\
\bottomrule
\end{tabular}
\label{tab:real_data_time}
\end{table}

\subsubsection{NIST Handwritten-Digit Dataset}
\label{sec:NIST_dataset}
The finite GMM is often used for model-based
clustering~\citep[Chapter 14.3]{friedman2001elements,fraley2002model}, which requires the learning of a mixture from the data.
When the dataset is large and/or distributed over local machines,
split-and-conquer approaches such as our proposed GMR
become useful. 
In this section, we demonstrate the use of the GMR method on the famous NIST dataset for character recognition~\citep{grother2016nist}.
We use the second edition, named
\emph{by\_class.zip}~\footnote{\scalebox{0.9}{Available at   \url{https://www.nist.gov/srd/nist-special-database-19}.}}. 
It consists of around 4M images of handwritten digits and characters (0--9, A--Z, and a--z) from different writers. 
This experiment focuses on the digits,
and we still refer to it as the NIST dataset. 
The images of the digits are in directories 30--39. 
According to the user guide\footnote{\scalebox{0.9}{Available at \url{https://s3.amazonaws.com/nist-srd/SD19/sd19_users_guide_edition_2.pdf}.}}, the images in the \emph{train\_30} to \emph{train\_39}
and \emph{hsf\_4} folders are used as the training and test sets respectively. The numbers of training images for each digit
are listed in the following table:
\begin{table}[!htpb]
\caption{The numbers of training images for each digit in NIST dataset.}
\centering
\small
\begin{tabular}{|c||cccccccccc|}
\hline
Digits & 0 & 1 & 2 & 3 & 4 & 5 & 6 & 7 & 8 & 9 \\
\hline
Training & 34803& 38049& 34184& 35293& 33432& 31067& 34079& 35796& 33884& 33720\\
Test & 5560& 6655& 5888& 5819& 5722& 5539& 5858& 6097& 5695& 5813 \\
\hline
\end{tabular}
\end{table}
Each image is a $128\times 128$ pixel grayscale matrix
whose entries are real values between $0$ and $1$ that
record the darkness of the corresponding pixels. 
A darker pixel has a value closer to $1$.
Following the common practice, we first trained a $5$-layer convolutional 
neural network and reduced each image to a $d=50$ feature vector of real values. 
The details of the neural network for the dimension reduction 
are given in Appendix~\ref{app:nn_architecture}. 
A na\"ive approach to building a classifier is to regard the features of each digit as a random sample from a distinct Gaussian. 
The pooled data is therefore a sample from a $10$-component finite GMM. 
We may learn this model based on the whole dataset or through split-and-conquer approaches.

\begin{figure}[htpb]
\centering
\subfloat[Training LL]{\includegraphics[width = 0.3\textwidth]{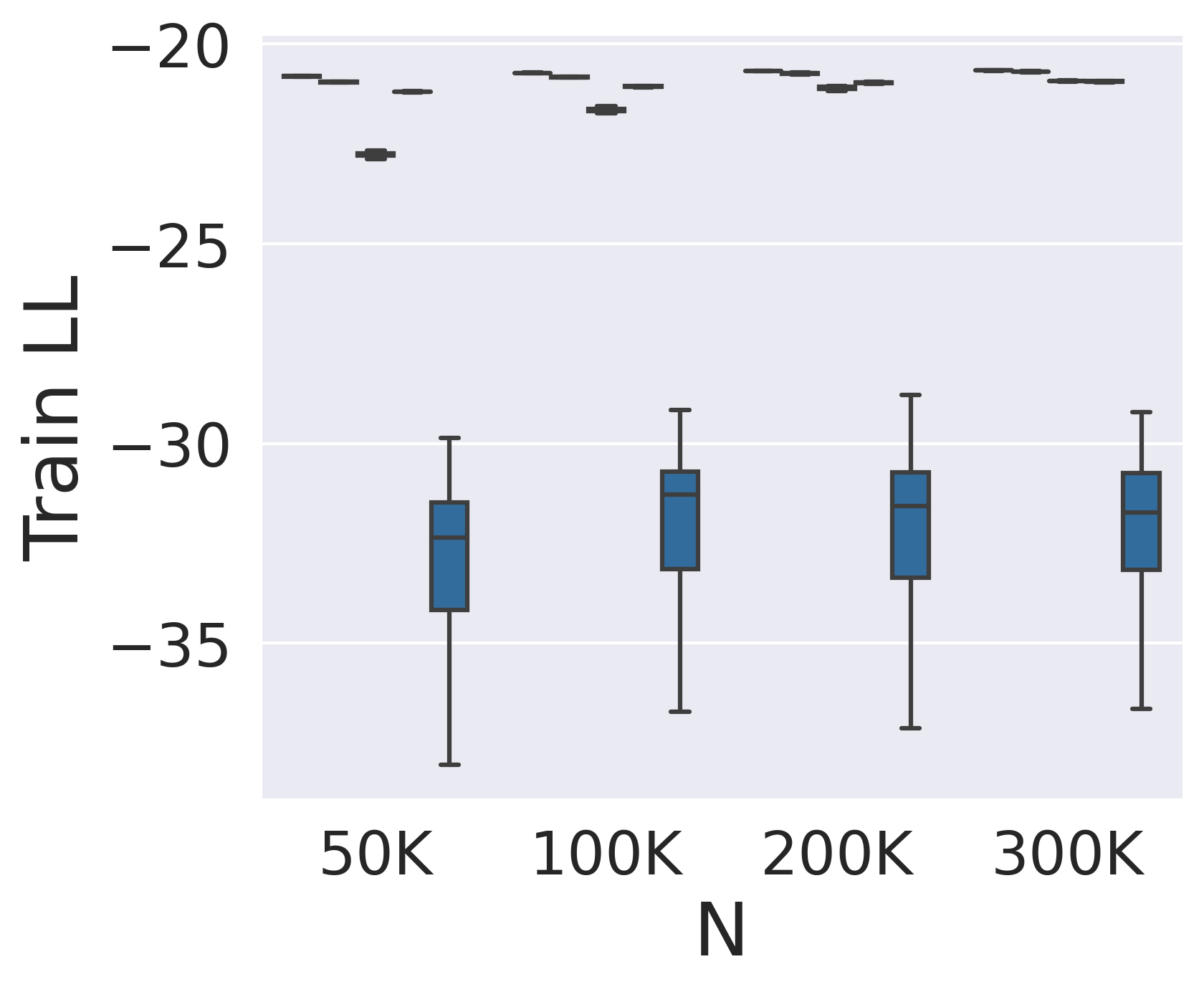}}
\subfloat[Test LL]{\includegraphics[width = 0.3\textwidth]{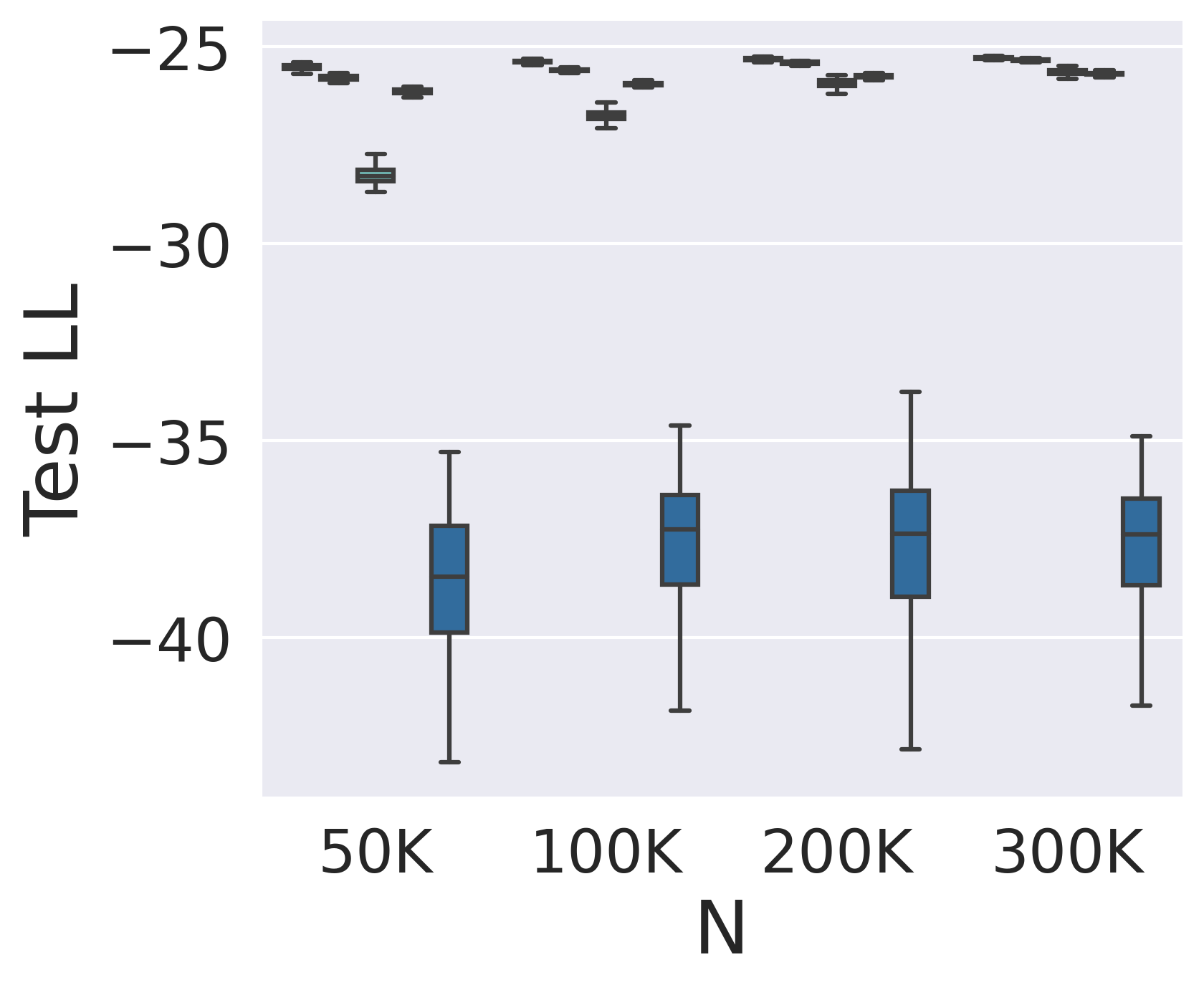}}
\subfloat[Computational time]{\includegraphics[width = 0.3\textwidth]{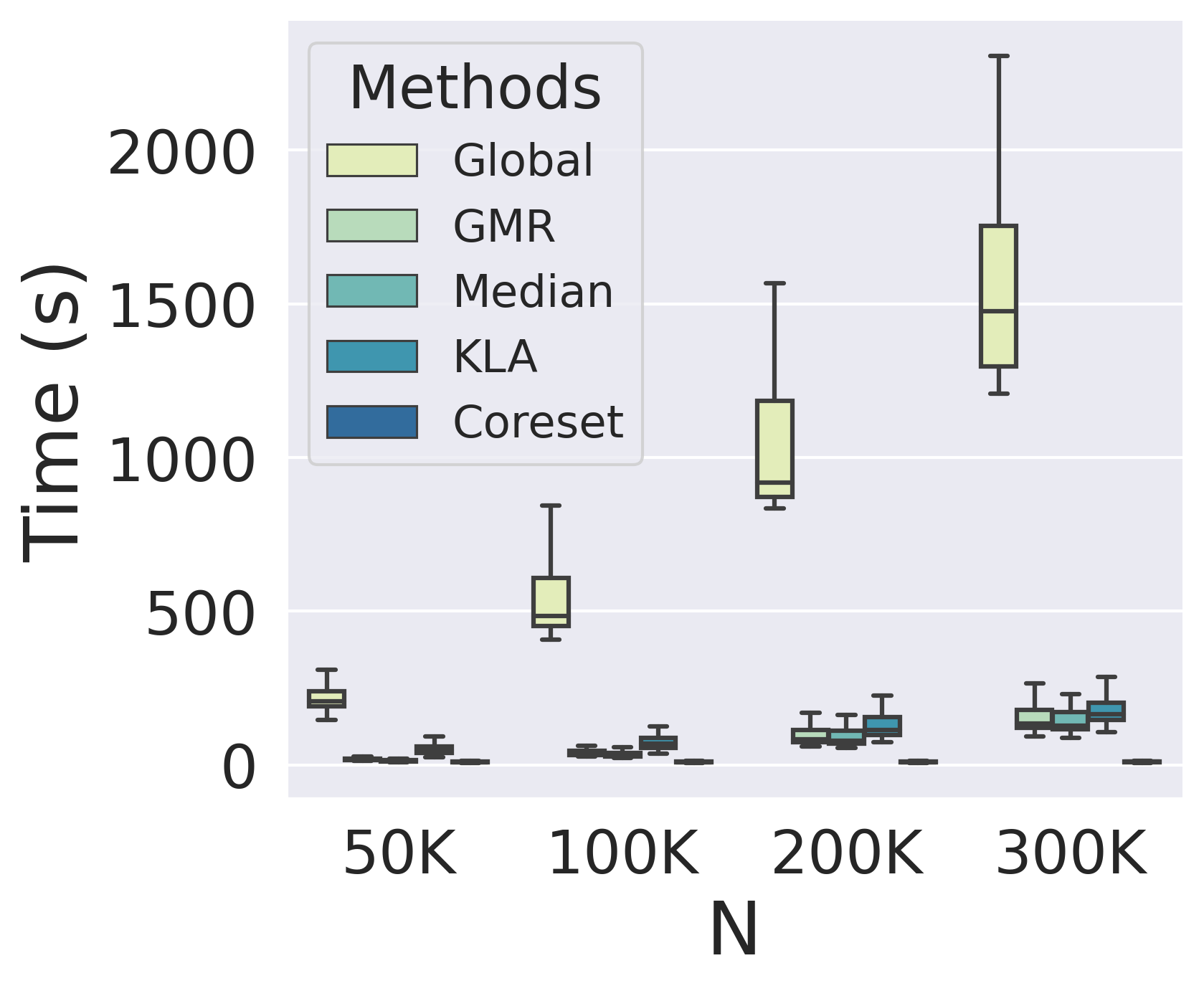}}\\
\subfloat[Training ARI]{\includegraphics[width = 0.3\textwidth]{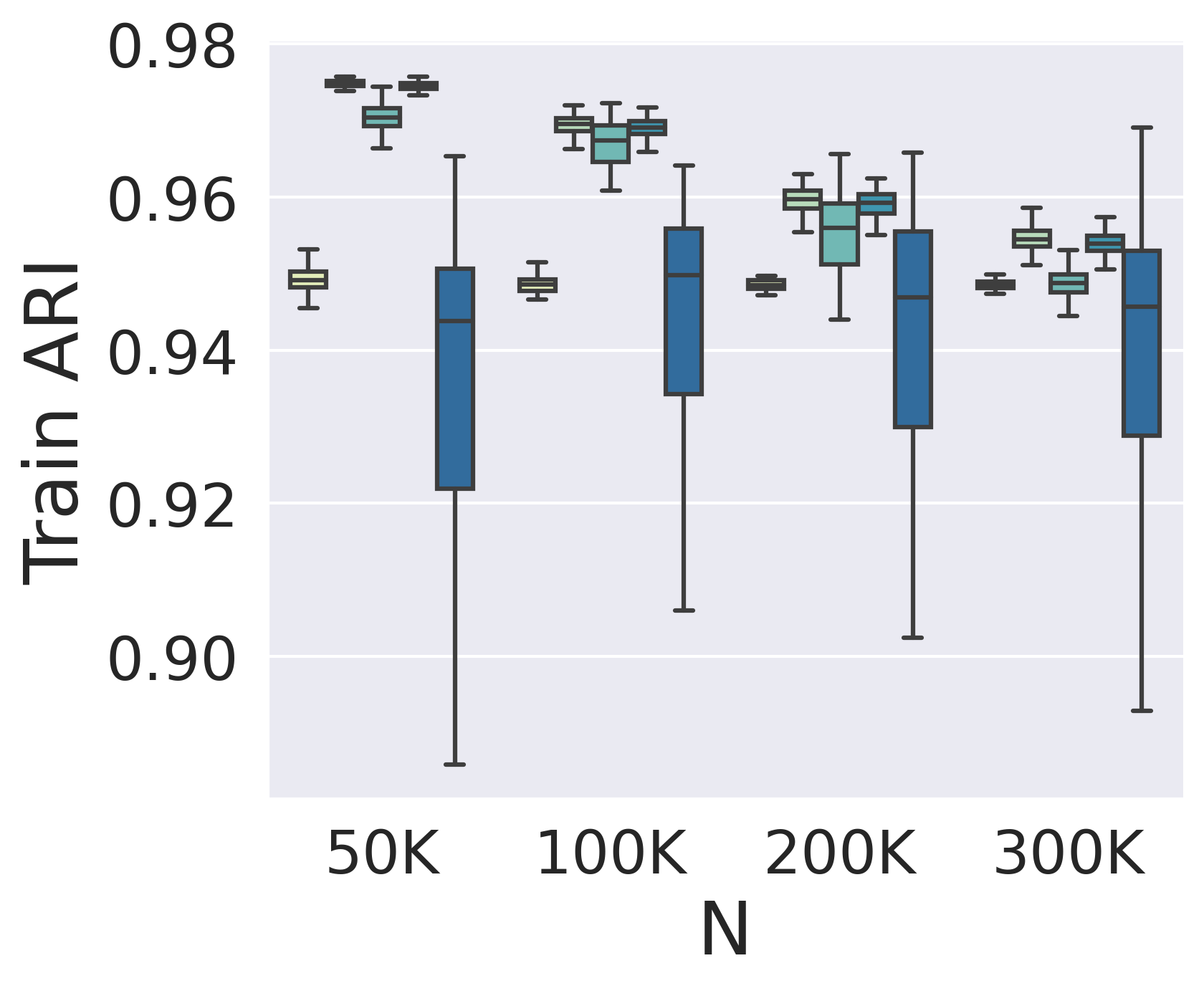}}
\subfloat[Test ARI]{\includegraphics[width = 0.3\textwidth]{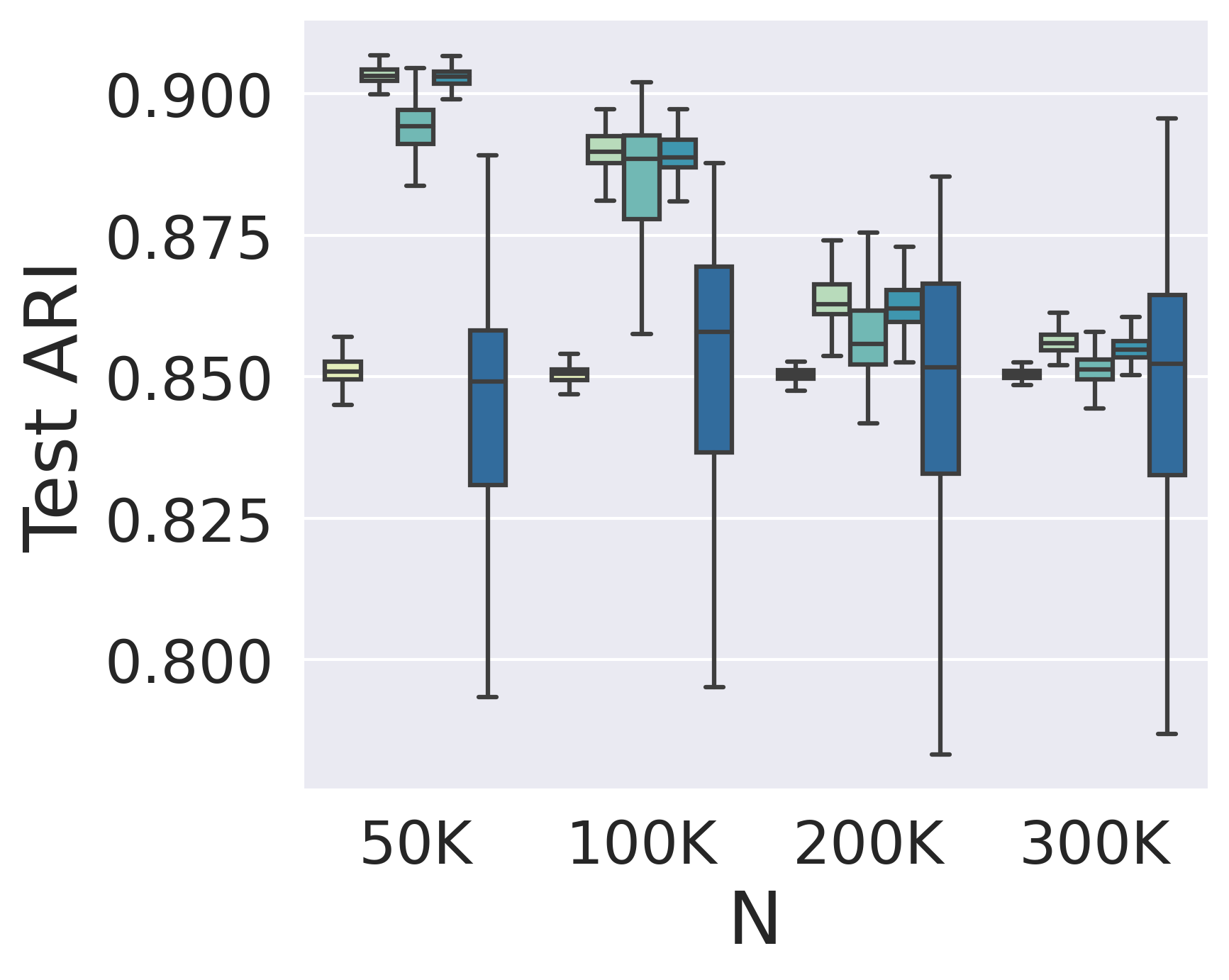}}
\caption{Performance of split-and-conquer approaches for learning of $10$-component Gaussian mixture on $50$-dimensional space for NIST digit classification.}
\label{fig:NIST}
\end{figure}

We randomly select $R=100$ datasets of size $N=50,000$ from the training set. 
Each dataset is then randomly partitioned into $M=10$ subsets.
We obtain global, GMR, Median, KLA, and Coreset estimates for a $10$-component GMM on each dataset.
The size of the generated sample for the KLA method and the coreset size in the Coreset method are both set to be $3000$.
This experiment was also carried out with the sample sizes $N=100K$, $200K$, and $300K$.
These mixture estimates are then used to cluster images of handwritten digits in the training and test sets. 
To assess the performance of different estimators, we compute the log-likelihood based on the training dataset and the test dataset.
To compare the performance of these estimators for clustering, we compute the ARI between the true label of the image and the predicted label based on~\eqref{eq:clustering}. 
Figures~\ref{fig:NIST}(a) and~\ref{fig:NIST}(b) respectively gives the boxplots of the log-likelihood based on the training set and the log-likelihood based on the test set.
The boxplots of the ARI on the training set and the test set are given in Figure~\ref{fig:NIST}(d) and Figure~\ref{fig:NIST}(e).

In terms of the likelihood value, our method gives the highest log-likelihood among all split-and-conquer approaches and the Coreset estimators is the worst. 
When the total sample size is small, the Median estimator is worse than the KLA estimator and their difference becomes smaller as the total sample size increases.
This is because the local sample size increases as the total sample size increases. 
When the total sample size is $300K$, the number of samples used to fit the local models and the samples generated to fit the aggregated model in the KLA approach are the same, it can be seen from the figure that the log-likelihood value of Median estimator and the KLA estimator are about the same.

In terms of the performance of the estimators for the purpose of clustering, surprisingly, the global estimator performs noticeably worse than the the split-and-conquer approaches although the global estimator has the highest log-likelihood value.
A likely explanation is that the 10-component finite GMM is a {\bf working} model rather than a {\bf true} model, whereas true models are used in simulated data. 
This eliminates the advantage of the global estimator. 
This can also be seen that an increased total sample size $N$ does not lead to an improved fit in general.
The ARIs get smaller as the total sample size increases, this is likely that we are more and more confident that the model is not accurate as more and more data comes in.  
Nevertheless, our method has the best performance in every case.
It has the highest average ARI values and smaller variations. 

Figure~\ref{fig:NIST}(c) shows that in terms of computational time, all split-and-conquer approaches saves the computational time. 
The Coreset estimator is the most efficient, the GMR and median estimators takes slightly longer and the KLA takes the longest time.

\subsubsection{Atmospheric Data Analysis}
\label{sec:CAM_dataset}
We follow~\citet{chen2013parallel} and apply our GMR approach
to fit a finite GMM to an atmospheric 
dataset\footnote{\scalebox{0.9}{Available at~\url{https://www.earthsystemgrid.org/dataset}.}}
named \emph{CCSM run cam5.1.amip.2d.001}.
These data were created by computer simulation based on Community Atmosphere 
Model version 5 (CAM5). 
The dataset contains daily observations of multiple atmospheric variables 
between the years 1979 and 2005 over 192 longitudes (lon), 
288 latitudes (lat), and 30 vertical altitude levels (lev). 
Our analysis only include variables:
the moisture content (Q), temperature (T), and vertical velocity ($\Omega$, OMEGA) of the air.

For ease of comparison, we analyze only observations 
in December, January, and February, i.e., winter in the northern hemisphere.
The number of days is thus 2430, and the restriction reduces the variation in the dataset. 
At each surface location, we filter out non-wet days (less than $1$\,mm of daily precipitation)
and focus on days with precipitation above the 95th wet-day percentile.
This step reduces the number of observations at each location,
not necessarily evenly. 
The analysis aims to cluster the locations according to
the multivariate variable of dimension $d=91$:
$30 \mbox{ lev} \times \mbox{ \{Q, T, $\Omega$\}}$ plus the daily
precipitation (PRECL) at the surface. 
Following~\citet{chen2013parallel}, we fit a finite GMM of order $K = 4$.
They suggest that this model is helpful in 
identifying modes of extreme precipitation in 3D atmospheric space 
over a few atmospheric variables.

After this preprocessing, the dataset still takes about 3\,GB of memory,
so we cannot learn a global mixture in a reasonable time.
We partition the dataset evenly into $M=128$ subsets
and apply our GMR approach with
the same numerical strategies as in the NIST experiments.
For comparison, we also aggregate the local estimates by the KLA
with $500$ observations generated from each local estimate.

\begin{figure}[!htp]
\centering
\includegraphics[width = 0.85\textwidth]
{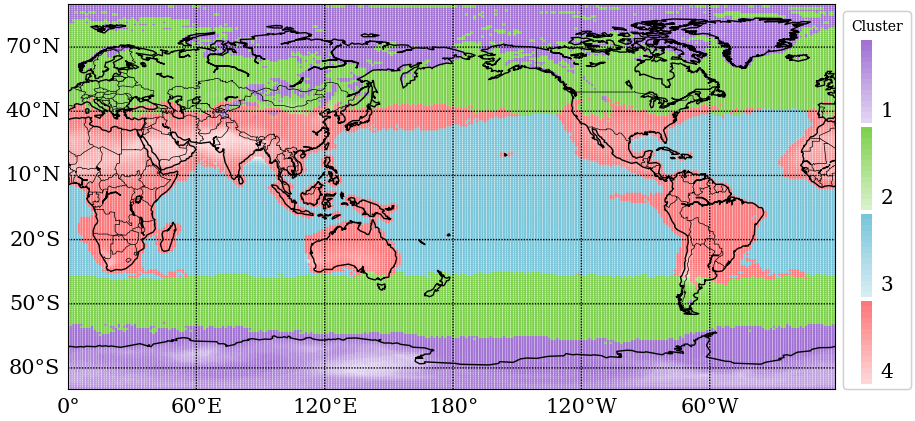}
\caption{Surface locations colored by clusters. Within each cluster, the darker the color, the more wet days at that location.}
\label{fig:CAM5}
\end{figure}

Once a finite GMM has been learned, we cluster the observations
based on~\eqref{eq:clustering}. 
Each combination of day and surface location is clustered into one of four subpopulations. To visualize the clusters, we further allocate each surface location to the cluster in which it belongs on most days. 
Figure~\ref{fig:CAM5} shows these four clusters represented by different colors.
Visually, the GMR clusters clearly separate the frigid, temperate, 
and tropical zones as well as the continental and oceanic areas. 
The KLA results in similar clusters.

\section{Discussion and Concluding Remarks}
\label{sec:conclusion}
We have developed an effective split-and-conquer approach for
the learning of finite GMMs. Our experiments show that it has
good performance both statistically and computationally.
We have focused on finite GMMs, but with some adjustment our approach
could be applied to learning mixtures with other
subpopulation distributions such as Gamma and Poisson.

We have ignored many potential issues.
In particular, we assume that the order of the
mixture is correctly specified and that
datasets on different local machines are IID and have
the same underlying distributions. Our method may be
regarded as a first step toward more satisfactory solutions to
these real-world problems.

\section*{Acknowledgement}
This research was enabled in part by support provided by WestGrid (\url{www.westgrid.ca}) and Compute Canada Calcul Canada (\url{www.computecanada.ca}).
The authors would like to thank Mario Lu\u ci\' c for providing the code in \citet{lucic2017training}.




{\small
\bibliography{biblio}
}
\clearpage
\section*{Appendices}
\input{supp}

\end{document}

%% file: mathcommands.tex
\usepackage{amsmath,amsfonts,bm}

\def\vx{{\bm{x}}}

\DeclareMathAlphabet{\mathsfit}{\encodingdefault}{\sfdefault}{m}{sl}
\SetMathAlphabet{\mathsfit}{bold}{\encodingdefault}{\sfdefault}{bx}{n}

\def\gC{{\mathcal{C}}}

\def\gX{{\mathcal{X}}}

\DeclareMathOperator*{\argmax}{arg\,max}
\DeclareMathOperator*{\argmin}{arg\,min}
\DeclareMathOperator*{\arginf}{arg\,inf}
\DeclareMathOperator*{\argsup}{arg\,sup}

\def\todo#1{\textbf{#1}}

\newcommand{\bmu}{\mbox{\boldmath $\mu$}}
\newcommand{\btheta}{\mbox{\boldmath $\theta$}}
\newcommand{\blambda}{\mbox{\boldmath $\lambda$}}

\newcommand{\bpi}{\mbox{\boldmath $\pi$}}

\newcommand{\bSigma}{\mbox{\boldmath $\Sigma$}}

\newcommand{\bv}{\mathbf{v}}
\newcommand{\bw}{\mathbf{w}}

\newcommand{\bbD}{\mbox{$\mathbb{D}$}}
\newcommand{\bbE}{\mbox{$\mathbb{E}$}}
\newcommand{\bbG}{\mbox{$\mathbb{G}$}}

\newcommand{\bbW}{\mbox{$\mathbb{W}$}}

\newcommand\raiseT[2]{%
  \setbox0\hbox{$#1{#2}$}\raise\dp0\box0}

\newcommand{\bea}{\begin{eqnarray*}}
\newcommand{\eea}{\end{eqnarray*}}
\newcommand{\ba}{\begin{eqnarray*}}
\newcommand{\ea}{\end{eqnarray*}}
\newcommand{\be}{\begin{equation}}
\newcommand{\ee}{\end{equation}}
\newcommand{\bi}{\begin{itemize}}
\newcommand{\ei}{\end{itemize}}

\newcommand{\bolds}{\boldsymbol}

\newcommand{\dKL}{D_{\text{KL}}}

\newcommand{\Xcal}{\mathcal{X}}
\newcommand{\Lcal}{\mathcal{L}}
\newcommand{\Tcal}{\mathcal{T}}

%% file: supp.tex
The appendix is organized as follows.
Section~\ref{app:proof} contains all left over technical details and proofs,
Section~\ref{app:algorithm} presents the pseudocode of our algorithm, and Section~\ref{app:additional_results} provides additional details.

\subsection{Proofs}
\label{app:proof}
\subsubsection{Example~\ref{eg:barycenter_of_GMM}: Technical Details}
\label{app:proof_barycenter_of_GMM}
We will show that  $\bbD(\bar G^C) < \bbD(\bar G)$ where
\[
\bbD(G) = 0.5 \bbW^2_{D, 2}(G_1, G) + 0.5 \bbW^2_{D, 2}(G_2, G).
\]
This result implies that $\bar G$ is not a barycenter. We stop
short of proving that $\bar G^C$ is. The latter task is so tedious
that we have it omitted.

Note that all transportation plans from $G_1$ and $G_2$ to
the presumed barycenter $\bar G^C$ have the form
\[
\begin{pmatrix}
p & 0.4 - p\\
0.4- p & 0.2 + p
\end{pmatrix}
\mbox{ and }
\begin{pmatrix}
p & 0.6- p\\
0.4- p & p
\end{pmatrix},
\]
respectively, for some $p$ between $0$ and $0.4$.
These two matrices are bivariate probability mass functions with the marginal probability masses 
$(0.4, 0.6)$ and $(0.6, 0.4)$ as required.
The cost functions may be presented as 
\[
\begin{pmatrix}
c(-1, -1)  &  c(-1, 2/3)\\
c( 1,  -1 )  & c( 1, 2/3) 
\end{pmatrix}
=
\begin{pmatrix}
0 &  25/9\\
4  & 1/9 
\end{pmatrix}.
\]
It is clear that $p = 0.4$ gives the optimal plans for transporting
 $G_1$ to $\bar G^C$ and $G_2$ to $\bar G^C$.
With these plans in place, we can see that
\[
\bbD(\bar G^C) = 0.5 \bbW^2_{D, 2}(G_1, \bar G^C) + 0.5 \bbW^2_{D, 2}(G_2, \bar G^C)
= 1/3.
\]
In comparison,  the optimal transportation plan 
from $G_1$ to $\bar G$ is to move $0.1$ mass from $\phi_1$ to $\phi_{-1}$
with a total cost of $0.1 \times 4 = 0.4$. Hence,
\[
\bbD(\bar G) = 0.5 \bbW^2_{D, 2}(G_1, \bar G) + 0.5 \bbW^2_{D, 2}(G_2, \bar G)
= 0.4 > 1/3.
\]
That is, $\bar G$ is not a barycenter.

\subsubsection{Proof of Theorem~\ref{thm:ww_averaging_equivalent_obj}}
\label{sec:app_equivalent_obj}

Let 
\(
G^* = \arginf \{\mathcal{J}_{c}(G): G \in \mathbb{G}_{K}\}.
\)
Let the mixing weights of any $G \in  \mathbb{G}_{K}$ be $\mathbf{v}(G)$,
and let the subpopulations prescribed by $G$ be $\Phi_{\gamma}$. According to~\eqref{eq:gmr_estimator_weight}, 
we have
\[
\bv (G^*) = \sum_i \pi_{i, \gamma}(G^{\star}),
\] 
which implies that $\bpi(G^*) \in \Pi(\cdot, \bv (G^*))$.
Since $\bpi(G^*) \in \Pi( \bw, \cdot)$ by~\eqref{Jcal}, we  also have that
$\bpi(G^{*}) \in \Pi(\bw, \bv (G^*) )$ or it is a valid transportation
plan from $\bar G$ to $G^*$.
Consequently,
\[
\inf\{\mathcal{T}_{c}( G): G \in \mathbb{G}_{K}\} 
\leq
\sum_{i, \gamma}  \pi_{i \gamma}(G^*) c( \Phi_i , \Phi^*_\gamma) 
= 
\mathcal{J}_{c}(G^*)
=
\inf\{\mathcal{J}_{c}(G): G \in \mathbb{G}_{K}\},
\]
with the last equality implied by the definition of $G^*$.
This inequality implies that the left-hand side of~\eqref{eq:equiv_optimization} is
less than its right-hand side.
    
Next, we prove that the inequality holds in the other direction.
Let $G^\dagger =\inf\{\mathcal{T}_{c}(G): G \in \mathbb{G}_{K}\}$, the solution
to the optimization on the left-hand side of~\eqref{eq:equiv_optimization}. 
We denote the subpopulations prescribed by $G^\dagger $ as $\Phi_\gamma^\dagger$.
Let 
\[
\bpi^\dagger = 
\arginf 
\big \{ 
\sum_{i, \gamma}  \bpi_{i \gamma} c( \Phi_i , \Phi^\dagger_\gamma): 
 \bpi \in \Pi(\bw, \bv(G^\dagger))
\big \},
\]
which is the optimal transportation plan from $\bar{G}$ to this $G^\dagger$. 
Because of this, we have
\[
\inf\{\mathcal{T}_{c}( G): G \in \mathbb{G}_{K}\} 
=
\Tcal_{c}( G^\dagger )
=
\sum_{i, \gamma}  \bpi^\dagger_{i \gamma} c( \Phi_i , \Phi^\dagger_\gamma) 
\geq
\mathcal{J}_{c}(G^\dagger)
\geq
\inf\{\mathcal{J}_{c}(G): G \in \mathbb{G}_{K}\}.
\]  
The last step holds because $\bpi^\dagger \in \Pi(\bw, \cdot)$.
This completes the proof.

\subsubsection{Proof of Theorem~\ref{thm:convergence}}
\label{app:condition_for_global_convergence_theorem}

{\bf (i)}. 
Clearly, we have
$\mathcal{K}_{c}(G|G_{t}) \geq \mathcal{J}_{c}(G)$
for all $G$ with equality holds at $G = G_t$.
Hence,
\begin{equation*}
\label{eq:MM_monotonicity}
\begin{split}
\mathcal{J}_{c}(G_{t})
&\geq
\mathcal{J}_{c}(G_{t}) - \{\mathcal{K}_{c}(G_{t+1}|G_{t}) - \mathcal{J}_{c}(G_{t+1})\}\\
&=
\mathcal{J}_{c}(G_{t+1})
 - \{\mathcal{K}_{c}(G_{t+1}|G_{t}) - \mathcal{J}_{c}(G_{t}) \}\\
 &\geq 
 \mathcal{J}_{c}(G_{t+1})
 - \{\mathcal{K}_{c}(G_{t}|G_{t}) - \mathcal{J}_{c}(G_{t}) \}\\
 &=
\mathcal{J}_{c}(G_{t+1}).
\end{split}
\end{equation*}
This is the property that an MM-algorithm must have.

\noindent
{\bf (ii)}.  Suppose $G^{(t)}$ has a convergent subsequence
leading to a limit $G^*$. Let this subsequence be $G^{(t_k)}$.
By Helly's selection theorem \citep{van2000asymptotic}, 
there is a subsequence $s_k$ of $t_k$ such that $G^{(s_k+1)}$ has a limit,
say $G^{**}$. 
These limits, however, could be subprobability
distributions. That is, we cannot rule out the
possibility/measure that the total probability in the limit is below 1 by
Helly's theorem.

This is not the case under the theorem conditions. Let $\Delta > 0$
be large enough such that
\[
A_1
=
\{ \Phi:  c(\Phi_i, \Phi) \leq \Delta, \mbox{for all subpopulations $\Phi_i$ of } \bar G\}
\]
is not empty. With this $\Delta$, we define
\[
A_2
= 
\{ \Phi:  c(\Phi_i, \Phi) > \Delta, \mbox{for all subpopulations $\Phi_i$ of  } \bar G\}.
\]
Suppose $G^\dagger$ has a subpopulation $\Phi^\dagger$ such that
$c(\Phi_i, \Phi^\dagger) > \Delta$ for all $i$.
Replacing this subpopulation in  $G^\dagger$
by any $\Phi^{\dagger \dagger} \in A_1$ to form $G^{\dagger\dagger}$, 
we can see that for any $t$,
\[
\mathcal{K}_c(G^\dagger | G^{(t-1)}) > \mathcal{K}_c(G^{\dagger\dagger}| G^{(t-1)}).
\]
This result shows that none of the subpopulations of $G^{(t)}$ 
are members of $A_2$.
Otherwise, $G^{(t)}$ does not minimize $\mathcal{K}_{c}(G|G^{(t-1)})$ at the $t$th iteration.

Note that the complement of $A_2$ is compact by condition \eqref{eq:compact.Phi}.
Consequently, the subpopulations of $G^{(t)}$ are confined to a
compact subset. Hence, all limit points of $G^{(t)}$, including both
$G^*$ and $G^{**}$, are proper distributions.
By the monotonicity of the iteration:
\[
\mathcal{J}_c(G^{(s_{k+1})}) 
\leq
\mathcal{J}_c(G^{(s_{k}+1)}) 
\leq
\mathcal{J}_c(G^{(s_{k})}).
\]
Let $k \to \infty$, we get 
\begin{equation}
\label{App.eq2}
\mathcal{J}_c(G^{**}) = \mathcal{J}_c(G^{*}).
\end{equation}

By the definition of the MM iteration, we have
$$
\mathcal{K}_{c}(G^{(s_{k}+1)}|G^{(s_k)})\leq \mathcal{K}_{c}(G|G^{(s_k)}).
$$
Let $k \to \infty$ and by the continuity of $\mathcal{K}_{c}(\cdot | \cdot)$, 
we get
\[
\mathcal{K}_{c}(G^{**}|G^{*})\leq \mathcal{K}_{c}(G|G^{*}).
\]
Hence, $G^{**}$ is a solution to $\min \mathcal{K}_{c}(G|G^{{t})})$ when $G^{(t)} = G^*$.
Namely,, we have $\mathcal{K}_{c}(G^{**}|G^{*}) = \mathcal{K}_{c}(G^{(t+1)}|G^{*})$.
With the help of \eqref{App.eq2}, it further implies
\[
\mathcal{J}_c(G^{**}) = \mathcal{J}_c(G^{(t+1)}) = \mathcal{J}_c(G^{*})
\]
when $G^{(t)} = G^*$. This shows that iteration from $G^{(t)} = G^*$ does not 
make $\mathcal{J}_c(G^{(t+1)})$ smaller than $\mathcal{J}_c(G^{(t)})$.
Hence, $G^*$ is a stationary point of the MM iteration.
This is conclusion (ii) and we have completed the proof.

%

\subsubsection{Proof of Theorem~\ref{GMR-consistency-KL}}
\label{app:consistency_proof}

Recall that the local estimators $\hat G_m, m \in [M]$ are strongly consistent
for $G^*$ when the order $K$ of $G^*$ is known.
Clearly, this implies that the aggregate estimator $\bar G \to G^*$
and $\Tcal_{c}(\bar G, G^*)  \to 0$ almost surely.
That is, other than a probability 0 event in the probability space $\Omega$
on which the random variables are defined,
convergence holds. 
Furthermore, each support point of $\bar G$ must 
converge to one of those of $G^*$.
The total weights of the support points of $\bar G$ converging to the
same support of $G^*$ must converge to the corresponding weight of $G^*$.
Without loss of generality,
assume $\Tcal_{c}(\bar G, G^*)  \to 0$ holds at all $\omega \in \Omega$
without a zero-probability exception.

By definition, $\bar G^R$ has $K$ support points. 
We also notice that
\be
\label{eq:fact}
\Tcal_{c}(\bar G, \bar G^R) \leq \Tcal_{c}(\bar G, G^*)  \to 0.
\ee
Suppose that $\bar G^R$ does not converge to $G^*$ at some $\omega \in \Omega$.
One possibility is that the smallest mixing weight of $\bar G^R$ 
(or a subsequence thereof) goes to zero as $N \to \infty$. 
In this case, $\bar G^R$ has $K-1$ or fewer meaningful
support points. 
Since the support points of $\bar G$ are in an infinitesimal neighborhood
of those of $G^*$, one of them must be a distance away from
any of the support points of $\bar G^R$. Therefore, by Condition 4, the transportation cost of this support point
is larger than a positive constant not depending on $N$.
The positive transportation cost implies that
$\Tcal_{c}(\bar G, \bar G^R) \not \to 0$, which contradicts
\eqref{eq:fact}.

The next possibility is that the smallest mixing weight of $\bar G^R$
does not go to zero. In this case, there is a subsequence such that
all the mixing weights converge to positive constants. 
Without loss of generality, all the mixing weights simply 
converge to positive constants as $N \to \infty$.
If there is a subsequence of support points of $\bar G^R$
that is at least $\epsilon$-distance away from any of the support
points of $G^*$, then the transportation cost from $\bar G$ to
this support point will be larger than a positive constant not 
depending on $N$. This again leads to a contradiction to
\eqref{eq:fact}.

The final possibility is that $\bar G^R$ (or a subsequence thereof)
has a proper limit, say $G^{**} \neq G^*$.
If so, $\Tcal_{c}(\bar G, \bar G^R) \to \Tcal_{c}(G^*, G^{**}) \neq 0$,
contradicting \eqref{eq:fact}.

We have exhausted all the possibilities. Hence, the consistency claim is true.

\subsubsection{Proof of Theorem~\ref{thm:rate_of_convergence}}
\label{app:convergence_rate_proof}

We start with a few rate conclusions.
Let $\Phi_{mk}$ be the $k$th  subpopulation learned at local machine $m$
and $w_{mk}$ be its mixing weight. Note that we do not put a ``hat'' on them
for notation simplicity.
According to Lemma~\ref{lemma:consistency_pMLE} on the rate of convergence of the
pMLE at local machines, these subpopulations can be
arranged so that for all $m\in[M]$ and $k\in[K]$, we have 
\bea
 \| \Phi_{mk} - \Phi^*_{k} \| = O_p(N^{-1/2}), ~~~
 \sum_m \lambda_m w_{mk} - w^*_k = O_p(N^{-1/2}).
\eea
By C5, the first rate conclusion above implies
\[
\max\{ c(\Phi_{mk}, \Phi^*_{k}): k \in [K]\} = O_p(N^{-1}).
\]
For each $k$, let $\widetilde w_k = \sum_{m=1}^M \lambda_m w_{mk}$
and $\widetilde \Phi_{k}$ be the local barycenter of $\Phi_{mk}$, $m \in [M]$:
\[
\widetilde \Phi_k = \argmin \{\Phi:  \sum_{m=1}^M \lambda_m w_{mk} c(\Phi_{mk},\Phi)\}.
\]
By the rate conclusions given earlier, 
we have $\widetilde w_k = w^*_k + O_p(N^{-1/2})$ for $k\in [K]$.
By C5,  we must also have 
\[
\| \widetilde \Phi_k - \Phi^*_k\| = O_p(N^{-1/2})
\]
and $\mathcal{T}_{c}(\bar G, \widetilde G) = o_p(N^{-1})$.
This  $\bar G^R$ is given by $\widetilde G$, then the rate conclusion
of the theorem is proved.

Next, we show that the GMR $\bar G^R$ is given by $\widetilde G$ asymptotically.
By theorem conditions, the true subpopulations $\Phi^*_k$ are all distinct. 
Hence, by condition C4, we have
\[
\min \{ c(\Phi^*_k, \Phi^*_{k'}): k \neq k' \in [K] \} > 0.
\]
Thus, if the subpopulations of $\bar G^R$ is not in an $o_p(1)$
neighbourhood of one of $\Phi^*_k$ even though everyone of $\bar G$ is, 
the transport cost to this subpopulation from any subpopopulation of $\bar G$
exceeds a positive constant in probability. This contradicts 
\be
\label{Appendix.eq1}
\mathcal{T}_{c}(\bar G, \bar G^R) \leq \mathcal{T}_{c}(\bar G, \widetilde G) = o_p(N^{-1}).
\ee
This implies, all subpopulations of $\bar G^R$ are within $o_p(1)$ neighbourhood 
of one of $\Phi^*_k$. Denote these subpopulations as $\bar \Phi_k^R$
The optimal plan must transport $\Phi_{mk}$ to $\bar \Phi_k^R$,
otherwise the total transport cost exceeds a positive constant in probability
which again contradicts \eqref{Appendix.eq1}.
Since $\bar G^R$ minimizes the transport cost, 
we must have $\bar \Phi_k^R = \widetilde \Phi_k$, the local barycenter.
These conclusions imply that GMR $\bar G^R=\widetilde G$ with probability 
approaching to 1. 
Consequently, the rates of convergence of $\widetilde w_k, \widetilde \Phi_k$
extend to those of $\bar G^R$ and this completes the proof.

\subsubsection{KL-divergence satisfies C5}
\label{app:kl_divergence_property}
Let $\bmu_1,\bSigma_1$ and $\bmu_2,\bSigma_2$ 
be the parameters of $\Phi_1$ and $\Phi_{2}$. 
It is known that
\[
 2 D_{\text{KL}}(\Phi_{1}\|\Phi_{2})
=
\log\{\text{det}(\bSigma_2)/\text{det}(\bSigma_1) \}
+ \text{tr}(\bSigma_2^{-1}\bSigma_1- \mathbf{I}_{d}) 
+ 
(\bmu_2- \bmu_1)^\tau \bSigma_2^{-1} (\bmu_2- \bmu_1).
\]
Assume both $\Phi_1$ and $\Phi_2$ are in a small neighborhood of $\Phi$ whose
covariance matrix $\bSigma$ is positive definite.
Hence, eigenvalues of $\bSigma_2$ are in small neighborhood
of these of $\bSigma$. Thus, there exists a positive constant $A_1$ such that
the second term in $ 2 D_{\text{KL}}(\Phi_{1}\|\Phi_{2})$ satisfies
\be
\label{App.eq4}
A_1^{-1}\|\bmu_2- \bmu_1\|^2 
\leq
 (\bmu_2- \bmu_1)^\tau \bSigma_2^{-1} (\bmu_2- \bmu_1)
\leq
A_1 \|\bmu_2- \bmu_1\|^2 
\ee
Let $\lambda_1, \ldots, \lambda_d$ be eigenvalues of  $\bSigma_2^{-1/2}\bSigma_1\bSigma_2^{-1/2}$. 
Since both  $\bSigma_1$ and  $\bSigma_2$ are in a small neighborhood of $\bSigma$, 
we have $\lambda_1, \ldots, \lambda_d$ all close to 1.
\[
\log \{\text{det}(\bSigma_2)/\text{det}(\bSigma_1) \}
+ \text{tr}(\bSigma_2^{-1}\bSigma_1- \mathbf{I}_{d}) 
=
\sum_{j=1}^d \{ (\lambda_j - 1) - \log \lambda_j \}.
\]
Note that $ (\lambda - 1) - \log \lambda$ is a convex function with its minimum
attained at $\lambda = 1$ at which point its second derivative equals 1.
Hence, there exists an $A_2 > 0$ such that 
\[
A_2^{-1} (\lambda - 1)^2 \leq (\lambda - 1) - \log \lambda \leq A_2 (\lambda - 1)^2.
\]
We have therefore shown that
\be
\label{App.eq5}
A_2^{-1}\sum_{j=1}^d (\lambda_j - 1)^2
\leq
\log \{\text{det}(\bSigma_2)/\text{det}(\bSigma_1)\}
+ \text{tr}(\bSigma_2^{-1}\bSigma_1- \mathbf{I}_{d}) 
\leq
A_2\sum_{j=1}^d (\lambda_j - 1)^2.
\ee
We now connect the bound with Frobenius norm.

For a positive definite matrix $\bSigma$, it is easy to see that $\| \bSigma-\mathbf{I}_{d}\|_F^2 = \sum_{j=1}^d (\sigma_j-1)^2$,
where $\sigma_1,\sigma_2,\ldots,\sigma_d$ are eigenvalues of $\bSigma$.
Frobenius norm also has sub-multiplicative property
\[
\| \bSigma_1 \bSigma_2\|_F \leq \| \bSigma_1\|_F \|\bSigma_2\|_F.
\]
Applying the sub-multiplicative property in our context, 
we get
\bea
\| \bSigma_1 -  \bSigma_2\|^2_F 
&\leq &
~ \| \bSigma_2^{1/2} \|^4_F ~
\| \bSigma_2^{-1/2}\bSigma_1\bSigma_2^{-1/2}- \mathbf{I}_{d}\|^2_F \\
&\leq &
A_3~ \| \bSigma_2^{-1/2}\bSigma_1\bSigma_2^{-1/2}- \mathbf{I}_{d}\|^2_F\\
&=&
A_3 ~\sum_{j=1}^d (\lambda_j - 1)^2
\eea
for some local positive constant $A_3 > \| \bSigma^{1/2} \|^4_F$, 
as both matrices are in a small neighborhood of $\Sigma$.
Similarly, we have
\bea
\sum_{j=1}^d (\lambda_j - 1)^2
&=&
 \| \bSigma_2^{-1/2} \bSigma_1 \bSigma_2^{-1/2}- \mathbf{I}_{d}\|^2_F
 \\
&=&
 \| \bSigma_2^{-1/2}\{ \bSigma_1 - \bSigma_2\} \bSigma_2^{-1/2}\|^2_F
 \\
 &\leq&
 \| \bSigma_2^{-1/2}\|^4_F ~ \| \bSigma_1 - \bSigma_2\|^2_F\\
&\leq&
A_4 \|  \bSigma_1 - \bSigma_2\|^2_F.
\eea
for some positive constant $A_4$.
This leads to
\[
\| \bSigma_1 - \bSigma_2\|^2_F
\geq 
A_4^{-1} \sum_{j=1}^d (\lambda_j - 1)^2.
\]
Let $A = 2 \max \{A_1, A_2, A_3, A_4\}$. Applying \eqref{App.eq4} and
\eqref{App.eq5},  we have
\[
A^{-1} \{ \| \bmu_1 - \bmu_2\|^2 + \| \bSigma_1 - \bSigma_2\|^2_F \}
\leq
  D_{\text{KL}}(\Phi_{1}\|\Phi_{2})
 \leq
A \{ \| \bmu_1 - \bmu_2\|^2 + \| \bSigma_1 - \bSigma_2\|^2_F \}
\]
when $\Phi_1, \Phi_2$ are in a small neighborhood of $\Phi$, with
$A$ being a positive constant depends on $\Phi$.
This shows that the KL-divergence has property C5.

\subsection{Algorithm}
\label{app:algorithm}
Algorithm~\ref{alg:mm_reduction} gives the pseudocode for the GMR estimator with the KL-divergence cost function.

\begin{algorithm}[htp]
\begin{algorithmic}
\State {\bfseries Initialization:} $\Phi_{\gamma}$, $\gamma\in [K]$
\Repeat
\For {$\gamma\in[K]$}
\For {$i\in[MK]$}
\State Let
\begin{equation*}
 \pi_{i\gamma}=
 \begin{cases}
 w_i & \text{if}~\gamma=\argmin_{\gamma'} \dKL(\Phi_{i}, \Phi_{\gamma'})\\
 0& \text{otherwise}
 \end{cases}
\end{equation*}
\EndFor
\State Let
\begin{equation*}
\begin{split}
\bmu_{\gamma} &=\sum_{i=1}^{MK} \{\pi_{i\gamma}/\pi_{\cdot\gamma}\} \bmu_{i}\\
\bSigma_{\gamma}&= \sum_{i=1}^{MK} \{\pi_{i\gamma}/\pi_{\cdot\gamma}\} 
\{
\bSigma_i + (\bmu_i-\bmu_{\gamma})(\bmu_i-\bmu_{\gamma})^\tau
\}  
\end{split}
\end{equation*}
\EndFor
\Until the value of the objective function $\sum_{i,\gamma}\pi_{i\gamma}\dKL(\Phi_{i}, \Phi_{\gamma})$ converges
\For {$\gamma\in[K]$}
\State Let $v_{\gamma} = \sum_{i} \pi_{i\gamma}$
\EndFor
\State {\bfseries Output:} $\{(v_{\gamma},\bmu_{\gamma}, \bSigma_{\gamma}): \gamma\in [K]\}$
\end{algorithmic}
\caption{MM algorithm for GMR estimator with KL-divergence cost function}
\label{alg:mm_reduction}
\end{algorithm}

\subsection{Additional Details}
\label{app:additional_results}
\subsubsection{Additional Simulation Results}
\label{app:more_simulation_results}
\begin{figure}[!ht]
\centering
{\includegraphics[width = 0.9\textwidth]{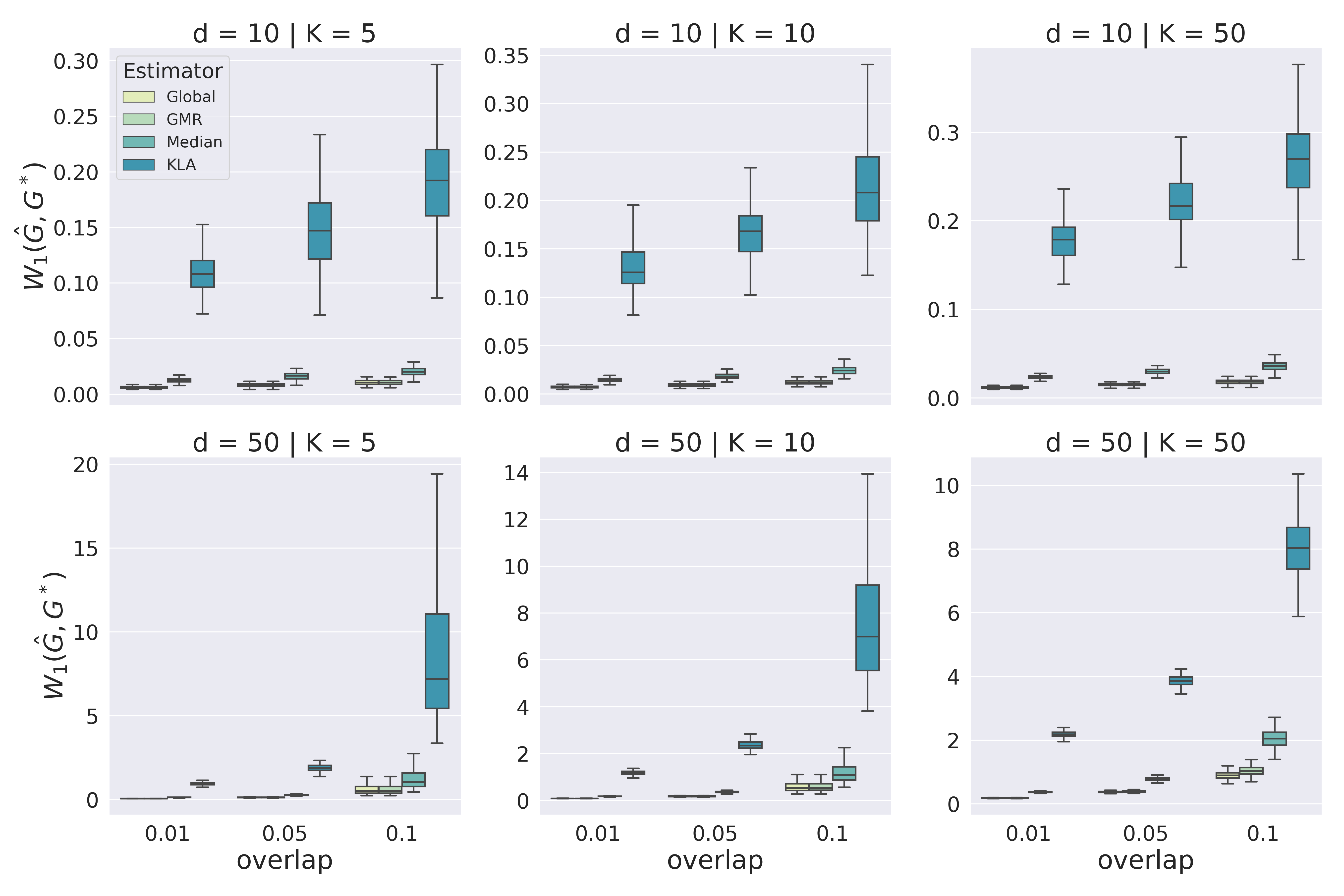}}
\caption{$W_1$ distances of estimators for learning mixtures.}
\label{fig:scalability_w1}
\end{figure}

\begin{figure}[!ht]
\centering
{\includegraphics[width = 0.9\textwidth]{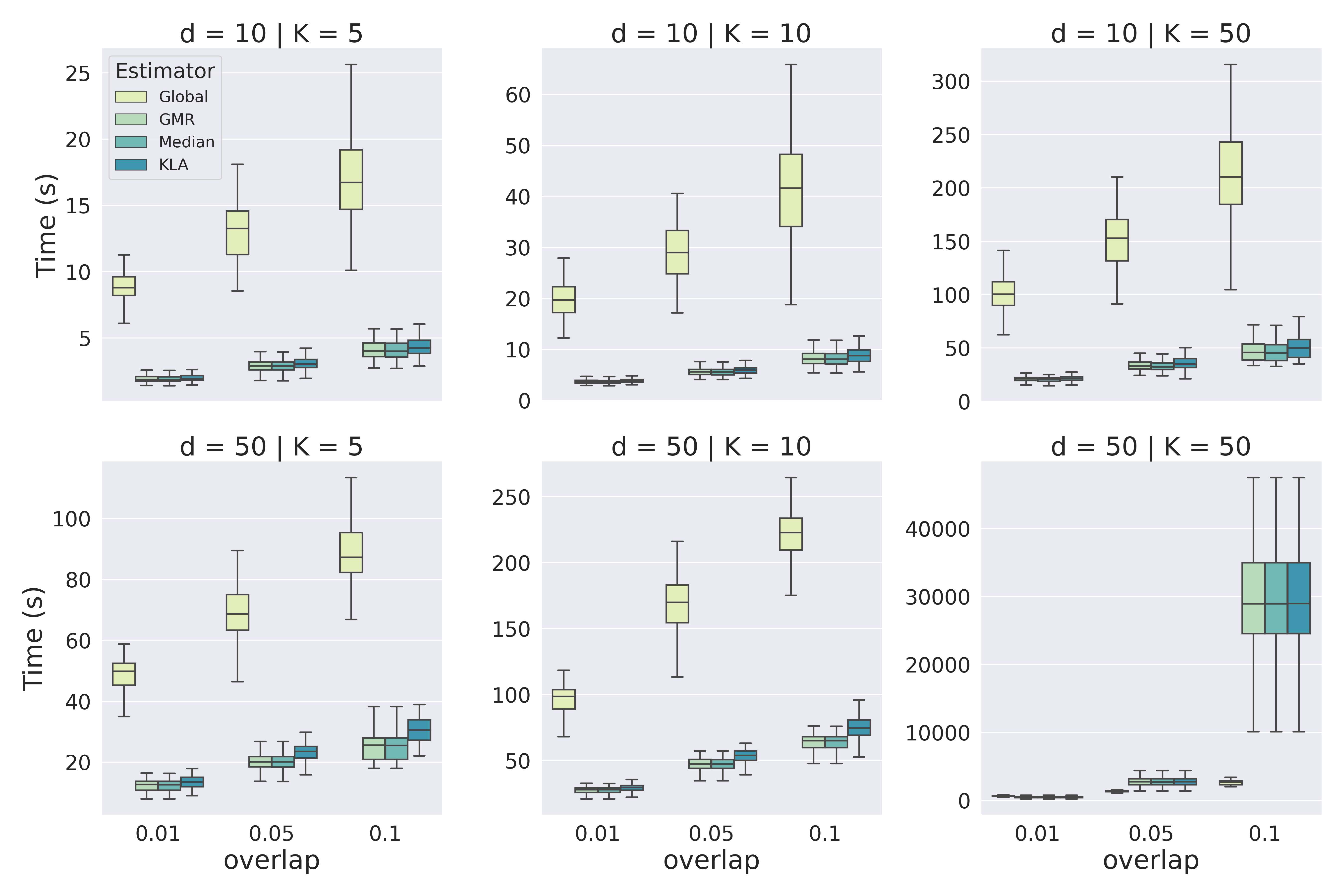}}
\caption{Computational times for learning mixtures.}
\label{fig:scalability_time}
\end{figure}

In this section, we present additional simulation results
for $K=5,10,50$ and $d=10,50$. 
All the settings are as in Section~\ref{sec:simulated_data}. 
Figures~\ref{fig:scalability_w1}--\ref{fig:scalability_time} show the
results for $N=2^{19}$ and $M=4$ with various combinations of $K$ and $d$. 
The panels in each figure are arranged so that the order of the mixture increases 
from left to right, and the dimension of the mixture increases from top to bottom. 

Comparing panels within the same row in Figure~\ref{fig:scalability_w1}, 
we note that the performance of all the estimators becomes worse as the order of 
the mixture increases in terms of $W_1$ distance. 
The panels within the same column in Figures~\ref{fig:scalability_w1} 
show that all the estimators become worse as the dimension of the mixture increases
in terms of both performance measures. 

Regardless, our estimator has performance
comparable to that of the global estimator. In terms of the misclassification error, 
for the same degree of overlapping,  the superiority of our 
estimator increases compared to the KL-averaging
as the number of components and the dimension increase.

The computational costs of the local estimators are typically low,
and this gives our method an added computational advantage.
However, this advantage is not guaranteed: see
the bottom right panel in Figure~\ref{fig:scalability_time},
where $d=50$, $K=50$, and the degree of overlapping is above $1\%$.
There are many other factors at play.
A more skillful implementation may lead to different conclusions
on the computational time.

\subsubsection{Convolutional Neural Network in NIST Example}
\label{app:nn_architecture}
Deep convolutional neural networks (CNNs) are 
commonly used to reduce the complex structure of a dataset to informative rectangle data. 
CNNs effectively perform dimension reduction and classification
in an end-to-end fashion~\citep{dara2018feature}. 
The final soft-max layer in a CNN can
be viewed as fitting a multinomial logistic regression model on the reduced feature space.
We use a CNN for dimension reduction in the NIST experiment; its
architecture is specified in Table~\ref{tab: architecture}. 
We implement the CNN in \emph{pytorch 1.5.0}~\citep{pytorch2019} 
and train it for $10$ epochs on the NIST training dataset. 
We use the SGD optimizer with learning rate $0.01$, momentum $0.9$, 
and batch size $64$. After the training, we drop the final layer 
and use the resulting CNN to reduce to 50 the dimension of the images 
in the training and test sets.

\begin{table}[!htp]
\caption{Architecture and layer specifications of CNN for dimension reduction in NIST example.}
\label{tab: architecture}
\centering
\begin{tabular}{c|c|c}
\toprule
 {\bf Layer} & {\bf Layer specification}&{\bf Activation function} \\
 \midrule
Conv2d & $C_{\text{in}}=1$, $C_{\text{out}}=20$, $H=W=5$&Relu\\
MaxPool2d & $k=2$ &--\\
Conv2d & $C_{\text{in}}=20$, $C_{\text{out}}=50$, $H=W=5$&Relu\\
MaxPool2d & $k=2$ &--\\
Flatten & -- &--\\
Linear & $H_{\text{in}}=800$, $H_{\text{out}}=50$&Relu \\
Linear & $H_{\text{in}}=50$, $H_{\text{out}}=10$&Softmax \\
\bottomrule
\end{tabular}
\end{table}